\documentclass[
aps,prd,   
               preprintnumbers,numbers,sort&compress,
               nofootinbib,
                            showpacs,
               colorlinks,
               linkcolor=blue,
               citecolor=blue]{revtex4-1}

   \newcommand{\exclude}[1]{}
\newcommand{\be}{\begin{eqnarray}}
\newcommand{\ee}{\end{eqnarray}}

\usepackage{graphicx,amsmath,amssymb,bm}

\usepackage{psfrag}
 \usepackage{feynmp}
 \usepackage{hyperref}

\usepackage{graphicx}
\usepackage{subcaption}

\newcommand{\beq}{\begin{equation}}
\newcommand{\eeq}{\end{equation}}

\def\ra{\rangle}
\def\la{\langle}

\begin{document}

\title{Gravitationally bound  axions  and how one  can discover  them} 
 
\author{Xunyu Liang}
  \email{xunyul@phas.ubc.ca}
 \author{Ariel  Zhitnitsky}
  \email{arz@physics.ubc.ca}
  
 \affiliation{Department of Physics \&  Astronomy, University of British Columbia, 
Vancouver,  Canada} 
   
 \begin{abstract}
 \exclude{We advocate the idea that  there  is a fundamentally new mechanism  for the axion production in the Sun and Earth  as recently suggested in \cite{Fischer:2018niu}.}
 As recently advocated in \cite{Fischer:2018niu}, there is a fundamentally new mechanism for the axion production in the Sun and Earth. However, the role of very  slow axions in previous  studies  were  neglected because of its negligible  contribution to the total  axion production by this new mechanism. 
  In the present work we specifically focus on analysis  of the 
  non-relativistic axions  which  will be trapped by the Sun and Earth due to the gravitational forces.  The corresponding emission rate of these low energy axions (below the escape velocity) is very tiny. However, these axions will be accumulated by the Sun and Earth during their  life-times, i.e. 4.5 billion of years, which  greatly enhances the discovery potential. 
     The computations  are based on the so-called Axion Quark Nugget (AQN) 
Dark Matter Model.    This    model was originally invented as a natural explanation of  the 
observed ratio $\Omega_{\rm dark} \sim   \Omega_{\rm visible}$ when the DM and visible matter densities assume the same order of magnitude  values, irrespectively to the axion mass $m_a$ or initial misalignment angle $\theta_0$.
 This model, without adjustment of any parameters,  gives a very reasonable intensity of the extreme UV  (EUV) radiation from the solar corona 
  as a result of the AQN annihilation events with the solar material.    This extra energy released in corona  represents a resolution, within AQN framework, a long standing puzzle known in the literature as the ``solar corona heating mystery".  The same annihilation events also produce the axions.  The flux of these axions is unambiguously fixed  in this model and expressed in terms of the   EUV luminosity from solar corona. 
  We make few comments on the potential    discovery of these gravitationally bound  axions. 
  \end{abstract}
  \maketitle

\section{Introduction}
The Peccei-Quinn (PQ) mechanism accompanied  by the axions remains the most compelling resolution of the strong CP problem, see original papers 
  \cite{axion,KSVZ,DFSZ} and  
 recent reviews  \cite{vanBibber:2006rb, Asztalos:2006kz,Sikivie:2008,Raffelt:2006cw,Sikivie:2009fv,Rosenberg:2015kxa,Marsh:2015xka,Graham:2015ouw,Ringwald:2016yge} on the subject. We refer to the review  articles for the discussions and analysis  on the  recent activities in the field of the axion searches   by a numerous number of different groups using very   different instruments.
 
 For the purposes of the present work it is sufficient  to mention  that the   conventional dark matter galactic  axions are  produced 
 due to the misalignment mechanism \cite{misalignment} when the cosmological field $\theta(t)$ oscillates and emits cold axions before it settles down at 
 its final destination $\theta_{\rm final}=0$.
 Another mechanism is due  to the   decay of the topological objects \cite{Chang:1998tb,Hiramatsu:2012gg,Kawasaki:2014sqa,Fleury:2015aca,Gorghetto:2018myk,Klaer:2017ond}.   There is a number of uncertainties and  remaining discrepancies in the corresponding estimates. We shall not comment on these subtleties\footnote{\label{DM}
  According to the most recent computations presented in ref.\cite{Klaer:2017ond}, the axion contribution to $\Omega_{\rm DM}$ as a result of decay of the  topological objects can saturate the observed DM density today if the axion mass is  in the range $m_a=(2.62\pm0.34)10^{-5} {\rm eV}$, while the earlier estimates suggest that the saturation occurs at a larger axion mass. One should also emphasize that the computations  \cite{Chang:1998tb,Hiramatsu:2012gg,Kawasaki:2014sqa,Fleury:2015aca,Gorghetto:2018myk,Klaer:2017ond} have been performed with assumption that PQ symmetry was broken after inflation.}
 by referring to the original   papers \cite{Chang:1998tb,Hiramatsu:2012gg,Kawasaki:2014sqa,Fleury:2015aca,Gorghetto:2018myk,Klaer:2017ond}. It is important that  in both cases
 the produced  axions are non-relativistic particles with typical $v_{\rm axion}/c\sim 10^{-3}$, and their contribution to the dark matter density scales as $\Omega_{\rm axion}\sim m_a^{-7/6}$. This scaling  unambiguously implies that the axion mass must be fine-tuned  $m_a\simeq 10^{-5}$ eV
 to  saturate the DM density today, see footnote {\ref{DM},  while larger axion mass will contribute very little to $\Omega_{\rm DM}$.
  The cavity type experiments have a potential to discover these  non-relativistic axions. 
 
 Axions can be also produced as a result of the Primakoff effect in a stellar plasma at high temperature \cite{Sikivie:1983ip}. These axions are ultra-relativistic as the typical average energy of the axions emitted by the Sun is $\la E\ra =4.2$ keV, see \cite{Andriamonje:2007ew}. 
 The searches for the solar axions are based on the helioscope  instrument   CAST  (CERN Axion Search Telescope) \cite{Andriamonje:2007ew}. 
 
It has been suggested in   recent  work \cite{Fischer:2018niu}
  that there is a fundamentally novel mechanism of the  axion production  in the Sun.  This mechanism is deeply rooted to the so-called axion quark nugget (AQN) dark matter model when the stability of the nuggets is supported by the axion domain wall. The most important  consequence of the new production mechanism is that 
  the emitted axions  (from the axion domain wall when the nugget gets annihilated)  will be released with relativistic  velocities  with 
typical values $v_{\rm axion}^{\rm AQN}\simeq 0.5 c$. These features should be   contrasted with conventional galactic non-relativistic axions    $v_{\rm axion}\sim 10^{-3}c$ and solar ultra-relativistic axions with typical energies 
$\la E\ra =4.2$ keV. 

The computations in ref.  \cite{Fischer:2018niu} of the spectral properties of the  axions produced by this novel mechanism  were based on the approximation which is known to be badly violated  for low-energy  part of the spectrum with $v\ll c$. This part of the spectrum represents very tiny 
portion of the produced axions. Therefore, it had been ignored in the original studies  \cite{Fischer:2018niu}. However, the upgraded CAST instrument will be highly sensitive to the low energy part of the spectrum. Therefore, it is highly desirable to develop a new computational technique   which would allow to carry out the computations  in the region of small velocities $v\ll c$. 

Furthermore, 
the low-energy axions produced in the Sun might be trapped by strong gravitational force such that $v\leq {v_{\rm trapped}}$ will be trapped by the Sun indefinitely. The $v_{\rm trapped}$ is  numerically the same as the free fall velocity,   
\be
  \label{trapped1}
   \frac{v_{\rm trapped}}{c}=\sqrt{\frac{2GM_{\odot}}{R_{\odot}}}\simeq 2\cdot 10^{-3}. 
  \ee
  While the portion of these low energy  axions is tiny as we shall estimate below, these trapped axions may play an important role in physics as they will be accumulated around the Sun during  entire life time of the solar system, i.e. around  4.5 billion years.
 The effects related   to  the trapped axions are  not new, and discussed previously in the literature \cite{LZ:2003}. The goal here is to present some numerical estimates for our specific AQN model when the axions
which are produced  as a result of the annihilation events   in the solar atmosphere and will be indefinitely  bounded to the Sun. 

Therefore, the main goal of the present studies is to develop a new technique to generalize  the results of ref.  \cite{Fischer:2018niu}   to perform the self-consistent  computations of the axion spectrum in the regime   when the axion velocities are small   $v\ll c$. 
 The corresponding generalization of the results \cite{Fischer:2018niu} requires to abandon   the so-called ``thin wall approximation" and develop some new technical tools which proper describe the regime with $v\ll c$.

 We should emphasize that the present work is a natural continuation of the previous studies \cite{Fischer:2018niu}. Therefore, to avoid repetition   we refer the readers to that paper with detail discussions of the AQN framework itself, its motivation, its consequences and predictions. The only comment we would like to make here is as follows.  
The AQN framework   is consistent with all known astrophysical, cosmological, satellite and ground based constraints. In fact, in a number of cases  some   observables become very close to present day constraints. 
 Furthermore, in few  cases the predictions of the model  may explain a number of the  long standing mysteries as overviewed  in  \cite{Fischer:2018niu}.
 
 The paper is organized as follows. 
 \exclude{In next section \ref{sec:QNDM} we overview the AQN model by paying special attention to the astrophysical and cosmological consequences of this specific dark matter  model. In section \ref{AQN-flares} we highlight  the basic arguments of ref.  \cite{Zhitnitsky:2017rop} advocating the idea that the annihilation events of the antinuggets with the solar material can be interpreted as the   nanoflares conjectured  by Parker      long ago.  The recent numerical simulations \cite{Raza:2018gpb} strongly support the original estimates  \cite{Zhitnitsky:2017rop} and explicitly show that the disintegration of the AQNs occurs precisely at the altitude  about 2000 km above the photosphere, in the vicinity of the so-called transition region (TR). Injection of this huge amount of energy  as a result of  the AQN annihilation  events    represents the resolution of the so-called `` the solar corona heating puzzle" within AQN framework.   The  disintegration of the nuggets inevitably  produces  the axions in the vicinity of the TR. The   computations of   the  intensity and spectral properties of these axions with relativistic velocities $v\sim c$  have been carried out  in   ref.\cite{Fischer:2018niu}.}
  In  Section \ref{spectrum} we develop a new technique which allows to generalize these computations 
 for low energy portion of the spectrum when $v\ll c$.  We use the corresponding results  in Section \ref{lensing} to discuss the physics of the trapped axions and we highlight the basic ideas   how to   discover them.    We conclude in Section \ref{conclusion} with few thoughts on the future developments.

\section{Axions from AQNs: Intensity and the Spectrum \label{spectrum}}

  The AQN model   was invented long ago  \cite{Zhitnitsky:2002qa} (though a specific formation mechanism   of the nuggets was 
 developed in  much more recent papers \cite{Liang:2016tqc,Ge:2017ttc,Ge:2017idw})
  as a natural explanation of  the observed ratio $ \Omega_{\rm dark}\sim \Omega_{\rm visible}$.  
  \exclude{The similarity between  dark matter $ \Omega_{\rm dark}$ and the visible matter $\Omega_{\rm visible}$  densities   strongly suggests that both types of matter  have been formed  during the same cosmological epoch, which must be the QCD transition as the baryon mass $m_p$ which represents the visible portion of the matter $\Omega_{\rm visible}$ is obviously  proportional to $ \Lambda_{\rm QCD}$, while the contribution related to the E\&W    physics proportional to the quark mass $\sim m_q$ represents only a
minor contribution to the proton mass.  
}
 In context of the present work  the argument supporting the AQN model goes as follows.  It has been known  for quite some time   that the  total intensity of the  observed EUV and soft x-ray radiation (averaged over time)   can be estimated as follows,
\be
\label{estimate}
   L_{\odot ~  (\rm from ~Corona)}  \sim 10^{30}\cdot \frac{\rm GeV}{\rm  s} \sim 10^{27}  \cdot  \frac{\rm erg}{\rm  s},
 \ee
 which represents (since 1939) the renowned  ``the solar corona heating puzzle".  The observations (\ref{estimate}) imply that the corona
 has the temperature  $T\simeq 10^6$K which is 100 times hotter than the surface temperature of the Sun, and conventional astrophysical sources fail to explain the EUV and soft x ray radiation from corona.

    It turns out that if one estimates   the extra energy being produced within the AQN dark matter scenario     one obtains
 the total extra energy $\sim 10^{27}{\rm erg}/{\rm  s}$    which 
precisely reproduces  (\ref{estimate})   for  the   observed EUV and soft x-ray intensities  \cite{Zhitnitsky:2017rop}.  The full scale Monte Carlo simulations \cite{Raza:2018gpb} support this  estimate.  One should add that the estimate   $\sim 10^{27}{\rm erg}/{\rm  s}$  for extra energy  is derived  exclusively in terms of known  dark matter density $\rho_{\rm DM} \sim 0.3~ {\rm GeV cm^{-3}}$ and dark matter  velocity $v_{\rm DM}\sim 10^{-3}c $ surrounding the sun  
 without adjusting any  parameters of the model.  We  interpret this ``numerical coincidence"  as an additional hint supporting the AQN model.  Our original remark relevant for the present work  is that if one accepts the explanation  that the solar corona heating puzzle is resolved within AQN scenario than the axion flux will be unambiguously  fixed in terms of the EUV observed luminosity (\ref{estimate}) as 
 the axion field  represents the crucial element in the AQN construction.

We start our presentation with  subsection \ref{intensity} where we highlight  the basic results  from ref.  \cite{Fischer:2018niu} by providing a self-contained text  for the convenience of the readers. In next subsections \ref{sec:4.2 Spectral properties} and \ref{Results} we explain the computational framework and present the results of the computations, referring to Appendix \ref{Appendix_3D} for the technical details. 

\subsection{Intensity}\label{intensity}
  The axions  play a key role in construction of the AQNs as they provide an additional pressure to stabilize the nuggets.   The corresponding axion contribution into the total nugget's energy density has been computed in \cite{Ge:2017idw}.   Depending on parameters the axion's contribution to the nugget's mass  represents about 1/3 of the total mass.
  It can be translated in terms of the axion luminosity from the Sun as   follows \cite{Fischer:2018niu}:
  \be
  \label{axions_rate}
  L_{\odot ~  \rm (axion)}\simeq   1.6 \cdot 10^{27} \cdot  \frac{\rm erg}{\rm  s}.
  \ee
   The corresponding  axion flux measured on Earth 
     can be computed  as follows  \cite{Fischer:2018niu}
  \be
  \label{axions}
  \Phi({\rm solar~axions}) \sim \frac{L_{\odot ~  \rm (axion)}}{4\pi \la E_a\ra D^2_{\odot}}\sim 0.3\cdot 10^{17}\frac{1}{\rm cm^2 ~s}
  \left(\frac{10^{-5} {\rm eV}}{m_a}\right), ~~~~~~~~ D_{\odot}\simeq 150\cdot 10^6 ~{\rm km},  
  \ee
  where we assume that the axion's energy  when the antinuggets get annihilated is slightly relativistic  $E_a\simeq 1.2 m_a$, but never becomes very relativistic. The corresponding energy flux is  \cite{Fischer:2018niu}
   \be
\label{energy-flux}
m_a  \Phi({\rm solar~axions}) \sim 3\cdot 10^{11}\frac{\rm eV}{\rm cm^2 ~s}.
\ee
These estimates  should be  compared  with conventional cold dark matter galactic axion contribution  assuming the axions  saturate the observed DM density:
\be
\label{galactic-axion}
m_a \Phi({\rm galactic ~axions})\sim  \rho_{\rm DM}\cdot v_{\rm DM}\simeq \frac{0.3~{\rm GeV}}{\rm cm^3}v_{\rm DM}\simeq 10^{16} \frac{eV}{\rm cm^2 ~s}. 
\ee

 Similar estimates can be also carried out for Earth. 
In this case as explained in  \cite{Fischer:2018niu}   the observations of the $E\&M$ showers due to the nuggets entering the Earth's atmosphere (before hitting the Earth's surface) require very large area detectors.  The nuggets will continue to radiate  $E\&M$ energy  in the 
deep underground. However, this radiation by obvious reasons cannot be recovered and observed. 
At the same time  the observation of the axions (which have been produced as a result of the annihilation events in the very deep underground) is possible, and in fact very promising. Indeed, 
the corresponding axion flux  can be estimated  as follows  \cite{Fischer:2018niu}    
\be
\label{earth-axion}
m_a \Phi ({\rm Earth ~axions}) \sim  10^{16}\cdot \left(\frac{\Delta B}{B}\right)\frac{\rm eV}{\rm cm^2 ~s}, 
\ee
   where  $\Delta B/B$ is the portion  of the AQNs  which get annihilated in the Earth's interior.        
   Interestingly, the axion flux  (\ref{earth-axion})   which is generated due to the
   AQN annihilation events is much larger than the  flux (\ref{energy-flux}) generated due to the AQN annihilation events in the solar corona and measured on Earth. 
   At the same time, the axion flux  (\ref{earth-axion}) is the same order of magnitude 
   as  the conventional cold dark matter galactic axion contribution (\ref{galactic-axion}). This is because  the parameter 
  ${\Delta B}/{B}\sim 1$ is expected to be order of one, as  a finite  portion of the AQNs  will get annihilated in the Earth's interior, which includes all components: the  crust, the mantle and the core.  However, the 
   wave lengths of  the axions produced due to AQN annihilations, are much shorter  due to their  relativistic velocities  $v\sim 0.5 c$, in contrast with conventional galactic isotropic axions with $v\sim 10^{-3}c$. Therefore, these two distinct contributions can be easily discriminated.

\subsection{Spectral properties. General Comments}
\label{sec:4.2 Spectral properties}
The basic idea of the computation of the spectrum is as follows. 
 Consider an AQN loosing its mass due to the annihilation with surrounding material, while  that the axion portion to the energy   remains  the same, as it is not linked to the annihilation processes. One should comment here that the axion domain wall in the equilibrium does not emit any axions as a result of pure kinematical constraint: the domain wall axions are off-shell axions in the equilibrium. The time dependent perturbation obviously changes this equilibrium configuration.  
In other words,  the  configuration becomes    unstable because the total energy of the system is no longer at its minimum. To retrieve its ground state, an AQN will therefore intend to lower its domain wall mass by radiating the  axions. To summarize: the emission of axions is an inevitable consequence during the annihilation of antinuggets in  simply for the reason to maintain the AQN stability.

Now, we want to identify a precise  mechanism which produces the on-shell freely propagating axions emitted by  the axion domain wall.   In this section we overview the basic idea   of the computational technique to be used. To address this question, we consider the general form of a domain wall solution:
\begin{equation}
\label{eq:4.2 phi soln}
\phi(R_0)=\phi_w(R_0)+\chi
\end{equation}
where $R_0$ is the radius of the AQN, $\phi_w$ is the classical solution of the domain wall, while  $\chi$ describes the excitations due to the time dependent perturbation.
We should note that, $\phi_w$ is clearly off-shell classical solution, while  $\chi$ describes the on-shell propagating axions.  Thus, whenever the domain wall is excited, namely $\chi\neq0$, freely propagating axions will be produced and emitted by the excitation modes. 

Few comments are in order before we proceed in subsection \ref{Results} with description of  the technical details    and corresponding results. First, if the domain wall can be considered to be infinitely large in $xy$ direction such that the profile function depends on a single variable $z$ the computations can be carried out easily  as the classical profile function $\phi_w(z)$ is known exactly. This is precisely the procedure which has been adopted in   previous paper  \cite{Fischer:2018niu}. The corresponding  technique is justified when a typical size $L_x\sim L_y \gg m_a^{-1}$ along $x,y$  is much larger than the width of the domain wall of order $m_a^{-1}$. If the wave length of the emitted axion is small, i.e. $\lambda_a \sim m_a^{-1}$ the axions  cannot carry  any information about the finite size of the system and the approximation is marginally justified  ($\lambda_a$ stands for the de Brogile wavelength of the emitted axion). This is precisely the approximation,  
the so- called ``thin wall approximation"  which has been adopted in computations  \cite{Fischer:2018niu}.  This approximate treatment is marginally justified for relativistic axions with $v\sim c$, and we expect that accounting for the finite size of the system  cannot drastically  change the results in the relativistic domain  $v\sim c$. This   will be explicitly confirmed below by present computations accounting for finite size of the system.

Secondly,  it is quiet obvious that  the  ``thin wall approximation"  is badly broken for non-relativistic axions with $v\ll c$ when $\lambda_a\gg m_a^{-1}$ and a new  technique must be developed  for proper analysis.  The basic idea of computation accounting for finite size of the system $R$ goes as follows. 
Suppose an AQN is traveling in vacuum where no annihilation event takes place, we expect the solution stays in its ground state $\phi(R_0)=\phi_w(R_0)$ which corresponds to the minimum energy state. Since there is no excitation (i.e. $\chi=0$), no free axion can be produced. However, the scenario   drastically changes when some baryon charge from the AQN get annihilated. Due to these annihilation processes, the AQN starts to loose a small amount of its mass, and consequentially its size shrinks from $R_0$ to a slightly smaller radius $R_{\rm new}=R_0-\Delta R$. Note that its quantum state $\phi(R_0)=\phi_w(R_0)$ is no longer the ground state, because a lower energy state $\phi_w(R_{\rm new})$ becomes available. Then, we may write the current state of the domain wall as $\phi(R_0)=\phi_w(R_{\rm new})+\phi_w'(R_{\rm new})\Delta R$, so the domain wall now has a nonzero exciting mode $\chi=\phi_w'(R_{\rm new})\Delta R$ and free axions can be produced during oscillations of the domain wall. To reiterate: the   annihilation of antinuggets with surrounding matter  forces the domain wall to oscillate. These   domain wall oscillations  generate excitation modes which  ultimately lead to radiation of the propagating axions.

Our last  comment deals with terminology and notations. The results for the spectrum obtained using the ``thin-wall approximation'' is coined as 1D spectrum. As we mentioned above  it  is marginally   justified when $\lambda_a \sim m_a^{-1}$, and it  admits mathematically exact treatment which was previously presented in \cite{Fischer:2018niu}.  In the present work we mostly deals with 3D computations when a finite size of the system plays a key role, which is always the case for $\lambda_a\gg  m_a^{-1}$.
 The potential pitfall is that  some     technical simplifications are required to treat the  3D case. Consequentially, the obtained  results might  be  sensitive to these technical simplifications.  In order to   characterize the sensitivity to our technical simplifications we introduce  a tunable parameter $\delta$ which varies from 0 to 1, so $\delta$ will serve as a probe to test   the sensitivity with respect to numerical simplifications. As we shall see below, the obtained  results are  not very sensitive to the choice of $\delta$.  Therefore we conclude that our 3D results are robust and reliable.

In what follows  we will express the normalized spectrum as a function of the speed of emitted axion $v_a/c$ defined in the nugget's frame, defined as follows
\be
\label{normalization}
 \rho(v_a)\equiv\frac{1}{\Phi^{\rm tot}_{\rm axions}}\frac{d}{dv_a}\Phi_{\rm axions} (v_a) , ~~
  ~~  ~~ \int^1_0 dv_a ~ \rho(v_a) =1~~~~, 
\ee
where $\Phi^{\rm tot}_{\rm axions}$ is the axion flux inserted to eq. (\ref{normalization}) for normalization purposes. It     assumes the magnitude $\Phi({\rm solar~axions})$ given by eq. (\ref{energy-flux}) for the solar axions, and the value $\Phi ({\rm Earth ~axions})$ given by  (\ref{earth-axion}) for the axions emitted from the Earth's core.

\subsection{Spectral properties. Results}
\label{Results}
We follow the procedure described above in subsection \ref{sec:4.2 Spectral properties} and present
the axion field in time dependent background  as follows
\be
\phi(t,r)=\phi_{w, \delta}(r-R_0)+\chi(t,r)
\ee
where $\phi_{w, \delta}(r-R_0)$ satisfies  the classical equation of motion while $\chi(t,r)$ describes the time-dependent excitations.
As exact solution  accounting for the finite size of the nugget is not known we parameterize different simplified solutions by parameter $\delta$. We consider parameter $\delta$ as a probe as explained above.
 
The next step   is to expand the action ${\cal S}[\phi]$ by keeping the quadratic terms only, 
\begin{equation}
\label{semiclassical}
{\cal S}[\phi]
={\cal S}[\phi_{w, \delta}]
+\int dt\int d^3x \left[
\frac{1}{2}\dot{\chi}^2-\frac{1}{2}\chi L_2[{\delta}]\chi
\right]+{\cal O}(\chi^3).
\end{equation}
 where $L_2[{\delta}]$ is the second order linear differential operator which depends on classical solution $\phi_{w, \delta}(r-R_0)$, and the parameter $\delta$ introduced here is a result of approximation to the true solution $\phi_{w}(r-R_0)$, see Appendix \ref{Appendix_3D} for the technical details.  The next step, as usual, is to expand the fluctuations $\chi$ in terms of complete basis and compute the coefficients $a_{plm}$ in this expansion. The result for the total radiated energy $E_{\rm rad}$  is given by eq.(\ref{eq:A E excitation}) from Appendix
   \ref{Appendix_3D}. It can be  presented it in the following form 
\be
\label{excitation}
E_{\rm rad}
=\int d^3x\frac{1}{2}\chi
\left[-\frac{\partial^2}{\partial t^2}+L_2[{\delta}]\right]\chi 
=\sum_{lm}\int d^3p\frac{1}{2}E_a |a_{plm}|^2  
=\sum_{lm}\int_{m_a}^\infty dE_a\cdot 2\pi p~E_a^2|a_{plm}|^2,  
\ee
 where the coefficients $a_{plm}$ can be explicitly computed and are given by (\ref{eq:A a_solution_2}).
The expression for the radiated energy (\ref{excitation}) allows us to compute the desired spectrum $ \rho(v_a)$ defined by (\ref{normalization}). The results of the computations are presented on Fig. \ref{fig:4.2 3D_rhov_del09} with three different choices of parameter: $\delta=0$, 0.5, and 1 for physically realistic conditions, see Appendix \ref{Appendix_3D} for the details.  The low energy portion of the spectrum with $0\leq v/c\leq 0.01$ is shown  on Fig.\ref{fig:4.2 3D_rhovE2_del00}.
  
 Few comments are in order. Parameter $\delta$ in our treatment of the problem  was introduced as a probe to test our computational scheme which requires to compute all the modes in the background of the classical solution parameterized by parameter $\delta$.  While the classical solution itself can be computed numerically, we need some analytical form 
 to proceed with computations of the modes. Parameter $\delta$ is precisely introduced in order to parametrize this analytical expression entering the differential operator  $L_2[{\delta}]$.  As mentioned above, the parameter  $\delta$ roughly varies from 0 to 1 in physically realistic circumstances.  With the purpose of the test we performed the  computations for different values of 
 $\delta$ shown in Figs. \ref{fig:4.2 3D_rhov}, where we also included the ``unphysical value" for parameter $\delta=8$ 
 exclusively for illustrative purposes. 
  
 \exclude{
 The corresponding  results are plotted in Appendix on Fig. \ref{fig:4.2 3D_rhov_del05} and Fig.  \ref{fig:4.2 3D_rhovE2_del05}
 for $\delta=0.5$ and Fig. \ref{fig:4.2 3D_rhov_del10} and Fig.  \ref{fig:4.2 3D_rhovE2_del10} for $\delta=1$.}
 One can 
 explicitly see that the results for the spectrum  are not very sensitive to parameter $\delta$.  As we discuss below,
 the crucial   factor $\xi$  to be introduced in next section and which enters all final formulae is also not sensitive to parameter $\delta$.  To reiterate:  the basic qualitative results are not very sensitive to  choice of parameter $0\leq \delta\leq 1$.  
  
  One next comment goes as follows. It is very instructive to compare our 3D computations with 1D computations presented in 
\cite{Fischer:2018niu}. We had anticipated  before the 3D computations have been carried out that the results   in the relativistic domain $v_a/c\gtrsim0.5$ should not be  drastically modified in comparison with simplified treatment in \cite{Fischer:2018niu}.
We can now confirm that this is indeed the case.  At the same time we had expected the drastic modification 
of the spectrum  in the  non-relativistic regime $v_a/c\leq 0.01$ which is the subject of the present work. Indeed, the    3D  spectrum in this domain behaves as $\rho (v_a)\sim v_a^3$ as shown in Fig. \ref{fig:4.2 3D_rhovE2_del00},
in contrast with linear dependence in  simplified treatment in ref. \cite{Fischer:2018niu}. This difference in behaviour at small $v_a/c\ll 1$ can be attributed to 
the phase  volume suppression $\sim d^3k$ in 3D case for $\lambda_a\gg m_a^{-1}$.

\begin{figure}[h]
    \centering
    \begin{subfigure}[b]{0.48\textwidth}
        \includegraphics[width=\textwidth]{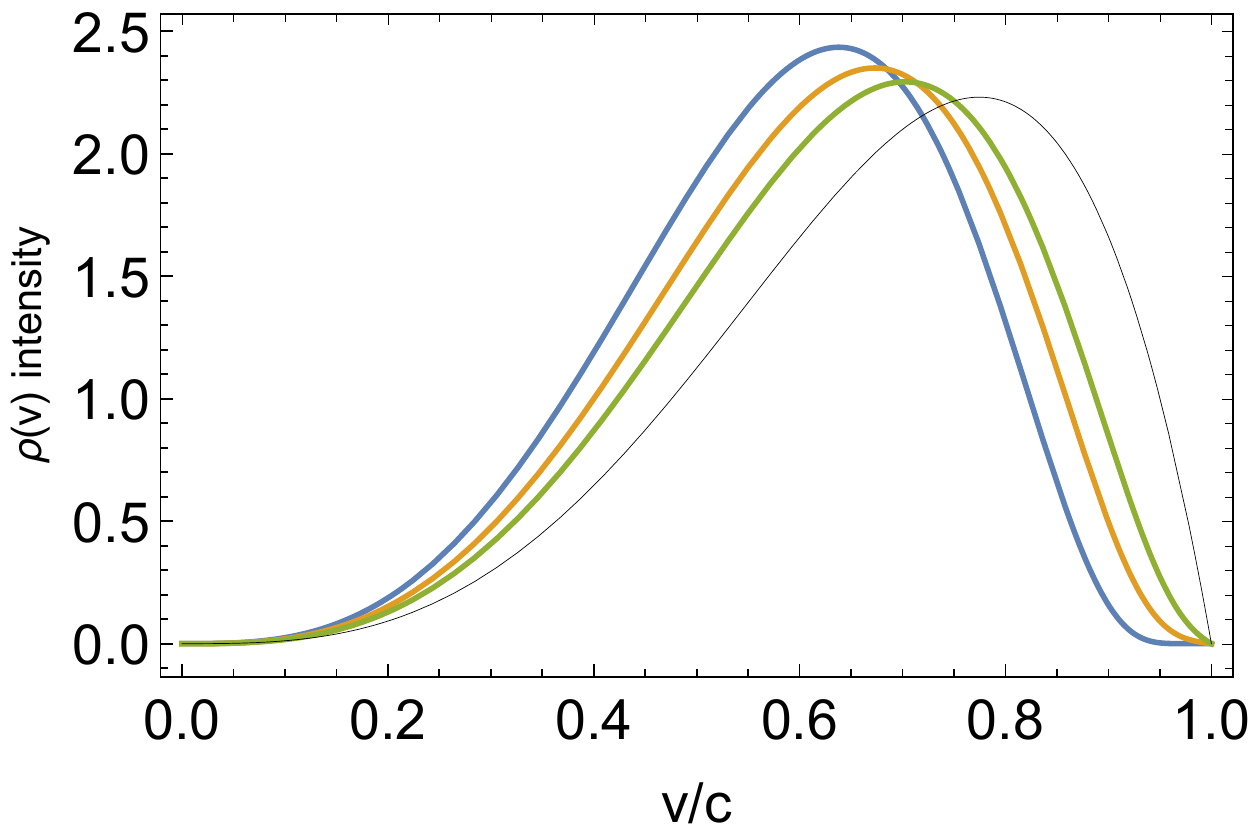}
        \caption{spectrum for  $0\leq v/c\leq 1$}
        \label{fig:4.2 3D_rhov_del09}
    \end{subfigure}
    ~ 
          \begin{subfigure}[b]{0.49\textwidth}
        \includegraphics[width=\textwidth]{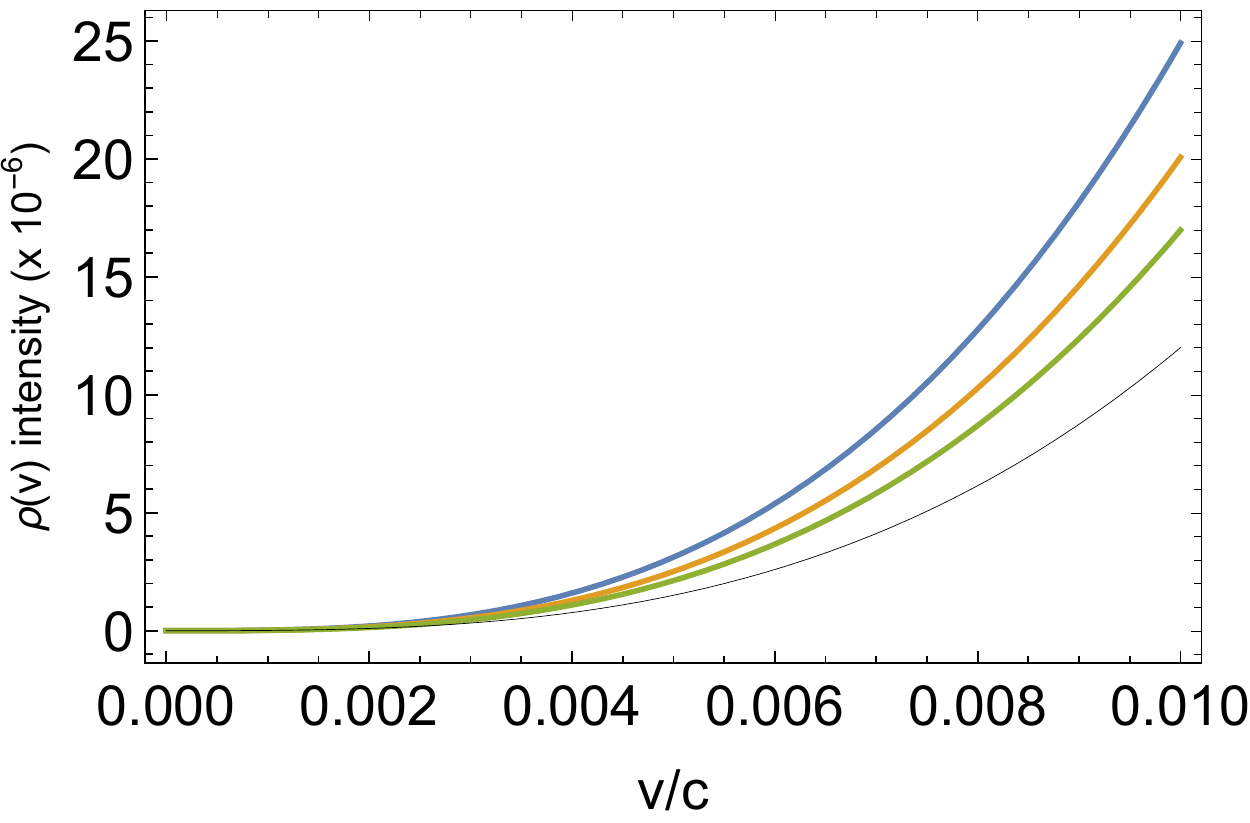}
        \caption{zoom in portion of the spectrum  with $0\leq v/c\leq 0.01$.}
        \label{fig:4.2 3D_rhovE2_del00}
    \end{subfigure}
     \caption{$\rho(v_a,\delta)$ vs $v_a/c$. Different values of $\delta$ are chosen respectively: 0 (blue), 0.5 (orange), 1 (green), and 8 (black).    }
     \label{fig:4.2 3D_rhov}
    \end{figure}

\exclude{
We also show the 1D result in Figs. \ref{fig:4.2 1D_rhov_del}. As expected, we see the 1D result gives qualitative consistent spectrum as in the 3D case especially in the relativistic domain $v_a/c\gtrsim0.5$. This is because in 1D case the thin-wall approximation is marginally justified in the relativistic domain (recall the condition $\lambda_a\lesssim m_a^{-1}$). On the other hand, the agreement is poor in the non-relativistic domain because the de Broglie wavelength $\lambda_a=1/m_av_a\gg m_a^{-1}$  is parametrically larger than the thickness of the domain wall  $\sim m_a^{-1}$. We note that the 3D case has a lot stronger suppression (by a factor of $10^{-5}$) comparing to the case in 1D.
}

\section{Gravitationally trapped axions}\label{lensing}

In the previous section we computed the portion of the axions  which   have sufficiently low velocities (below escape velocity)
such that they  will be orbiting the Sun as long as it exists. This portion of the non-relativistic axions is extremely tiny as shown on Fig. \ref{fig:4.2 3D_rhovE2_del00}. Nevertheless, the effect could be drastically enhanced as we discuss below due to  accumulation of these axions during entire life of the solar system, i.e. for $\sim $4.5 billion of years. 

The condition for the axions to be bounded  after being radiated  is determined by the trapped velocity ${v_{\rm trapped}}$, 
defined as 
\be
  \label{trapped}
   \frac{v^{\rm trapped}_{\odot}}{c}=\sqrt{\frac{2GM_{\odot}}{c^2R_{\odot}}}\simeq 2\cdot 10^{-3},  ~~~~~ 
   \frac{v^{\rm trapped}_{\oplus}}{c}=\sqrt{\frac{2GM_{\oplus}}{c^2R_{\oplus}}}\simeq 3.7\cdot 10^{-5}, 
   \ee
   such that all axions with $v\leq {v^{\rm trapped}_{\odot}}$ will be trapped by the Sun and the axions with $v\leq {v^{\rm trapped}_{\oplus}}$
   will be trapped by the Earth. The effect of the trapped axions is not new, and discussed previously in the literature \cite{LZ:2003}. The goal here is to present some numerical estimates for our specific AQN model when the axions
which are produced  as a result of the annihilation events  can  be trapped in the solar atmosphere. These estimates will play a key role in  our discussions on the discovery potential of these axions. 
\subsection{Solar corona background. Non-resonance case. \label{non-resonance}}
According to Fig. \ref{fig:4.2 3D_rhovE2_del00}
these highly non-relativistic axions represent a very tiny portion of the    produced axions. The energy which is accumulated in the solar atmosphere per unit time as a result of trapping   these axions  can be estimated  as follows
 \be
  \label{axions_trapped}
  \frac{d E_{\odot}}{dt} ({\rm trapped ~axions})\simeq    1.6 \cdot 10^{27} \cdot  \xi \cdot \frac{\rm erg}{\rm  s} \simeq 10^{17} \cdot  \left(\frac{\xi}{10^{-10}}\right) \cdot  \frac{\rm erg}{\rm  s}
  \ee
  where we  used the expression  (\ref{axions_rate}) for the rate of the energy transfer to the axions. We also     introduced the suppression factor $\xi$ to account for the small fraction of the trapped axions with $v\leq {v_{\rm trapped}}$. For numerical estimates  in formula  (\ref{axions_trapped})   we use   suppression factor $\xi \sim 10^{-10}$ computed\footnote{To demonstrate  the insensitivity to parameter  $\delta$, we note that $\xi$ shows very moderate   changes  between  $(0.68-1)\times10^{-10}$ when  $\delta$ varies between  $0$ and $1$. For the ``unphysical value" $\delta=8$ the parameter  $\xi\simeq0.48\times10^{-10}$, see Appendix \ref{Appendix_3D}.}
  in previous section 
  and presented on Fig. \ref{fig:4.2 3D_rhovE2_del00}.

    The axions  (\ref{axions_trapped}) could not  leave the system  during   entire life time of the Sun, i.e. 4.5 billion years $\simeq 10^{17}{\rm s}$. Therefore, the total energy accumulated  by the Sun and related to AQN annihilation events   radiating  the slow velocity  axions can be estimated as follows
  \be
  \label{axions_total}
   E_{\odot}  ({\rm trapped ~axions})  \simeq 10^{17} \cdot  \left(\frac{\xi}{10^{-10}}\right) \cdot  \frac{\rm erg}{\rm  s}\cdot 10^{17}s\simeq 10^{34}  \left(\frac{\xi}{10^{-10}}\right){\rm erg}.
  \ee
  This energy can be expressed  in terms of extra solar mass $\Delta M_{\odot}$  accumulated  by the Sun and represented by the trapped axions
   \be
  \label{axions_mass}
 \Delta M_{\odot}   ({\rm trapped ~axions})  \simeq 10^{10}   \left(\frac{\xi}{10^{-10}}\right){\rm kg},
  \ee 
  which of course represents a very tiny fraction of the solar mass $ M_{\odot}\simeq 2\cdot 10^{30} {\rm kg}$.
  
The energy (\ref{axions_total}) corresponds to the following total number of the axions accumulated  by the Sun during  its life-time:
 \be
  \label{number_total}
   N_{\odot}^{\rm axions} \sim \frac{E_{\odot}  ({\rm trapped ~axions}) }{m_ac^2} \simeq 10^{51} \cdot  \left(\frac{\xi}{10^{-10}}\right) \cdot \left(\frac{10^{-5} {\rm eV}}{m_a}\right).
  \ee
If we assume that the majority of these axions are localized within 2 solar radius $R_{\odot}$, we arrive to the following estimate for the average axion energy density inside this volume
\be
  \label{energy_density}
 \rho_{\odot}^{\rm axions}\sim \frac{E_{\odot}  ({\rm trapped ~axions})}{\frac{4}{3}\pi (2R_{\odot})^3} \sim   0.5\cdot  10^{3}    \left(\frac{\xi}{10^{-10}}\right)   \frac{\rm GeV}{\rm cm^3}, 
  \ee
which is 3 orders of magnitude larger than the present average dark matter density today $\rho_{\rm DM}\simeq 0.3 ~ \frac{\rm GeV}{\rm cm^3} $. One should comment here that this enhancement of the DM density in the vicinity of the Sun obviously not in contradiction with most precise observational upper limits on solar system (SS) -bound DM, which is normally expressed as 
$\rho_{\rm SS}< 2\cdot 10^5 \frac{\rm GeV}{\rm cm^3}$,
see e.g.\cite{Adler:2008vc}. It is also interesting to note that some authors \cite{Xu,Khriplovich} previously argued that the DM in the SS might be greatly enhanced (on the level of $10^3$) as a result of capturing of DM particles from the Galactic halo due to the 3 body interaction (the Sun, a planet and DM particle).  Other authors \cite{Edsjo,Peter} estimated that the effect of capturing is small. We refer to these original papers for the discussions and details. The only comment we would like to make here is that the effect  estimated  in eq. (\ref{energy_density}) is fundamentally distinct  in nature in comparison with previously discussed effect \cite{Xu,Khriplovich,Edsjo,Peter}. The novel effect which is the subject of this work is entirely rooted to the AQN model when the nuggets get disintegrated when enter the solar atmosphere. The corresponding annihilation events produce the  low velocities axions with  $v\leq {v^{\rm trapped}_{\odot}}$. These axions which behave as DM particles surrounding the Sun have no relation to the    effect discussed in 
\cite{Xu,Khriplovich,Edsjo,Peter}.

Now we want to estimate the number density $ n_{\odot}^{\rm axions}$ of these axions assuming, as before,  that the majority of the axions are localized within 2 solar radius $R_{\odot}$. 
 \be
  \label{number_density}
   n_{\odot}^{\rm axions}=    \frac{N_{\odot}^{\rm axions}  }{\frac{4}{3}\pi (2R_{\odot})^3} \simeq 0.5\cdot  10^{17}   \left(\frac{\xi}{10^{-10}}\right) \cdot \left(\frac{10^{-5} {\rm eV}}{m_a}\right) \frac{1}{\rm cm^3} .
  \ee

Can these axions be observed? These axions cannot decay as the axion life time $\tau(a\rightarrow 2\gamma)$  is very long. However, these axions can be converted to photons in the  background  of external magnetic field. The corresponding probability of this conversion is determined by the   formula  \cite{Raffelt:1987im,Ioannisian:2017srr}:
\be
P_{a\rightarrow \gamma}=\sum_{q=q_{\pm}}\left(\frac{g_{a\gamma}\cal{B}}{q}\right)^2\sin^2\left(\frac{qL}{2}\right), ~~~~~~ q_{\pm}=\pm \omega-\sqrt{\omega^2-m_a^2} 
\label{conversion}
\ee
where $L$ is a typical distance where the magnetic field $\cal{B}$ is  present. For non-relativistic axions one can approximate $q_{\pm}\simeq \pm \omega$.
 Furthermore, for our present analysis we assume that typical  ${\cal{B}}\sim 300$ G in  
the solar atmosphere, while $L$ is very large\footnote{A rough estimate in the following subsection suggests that $Lm_a\sim10^{3}$ if the resonance condition   is satisfied, see Eq.  (\ref{enhancement}). 
For more general non-resonant case a typical length scale is even   larger  within the classical axion windows, $10^{-6}{\rm eV}<m_a<10^{-3}{\rm eV}$ because there is no requirement for the variation of the plasma frequency on scale $L$ to be small.} such that  $\sin^2\left(\frac{qL}{2}\right)$ can be approximated as $\frac{1}{2}$. Therefore, probability of the conversion can be approximated as follows
\be
P_{a\rightarrow \gamma}\simeq \left(\frac{g_{a\gamma}{\cal{B}}}{m_a}\right)^2 ~~~~~~~{\rm where} ~~~~  \frac{g_{a\gamma}}{m_a}\simeq \frac{\alpha}{2\pi (m_{\pi}f_{\pi})}\cdot\left (\frac{E}{N}-\frac{2}{3}\frac{4+z}{1+z}\right)\frac{1+z}{\sqrt{z}}, 
\label{conversion1}
\ee
where $z=m_u/m_d\simeq 0.56$ and parameter $E/N=0$ for KSVZ model, and  $E/N=8/3$ for DFSZ model. For simple numerical analysis   
we take $E/N=0$ to arrive to the following estimate 
\be
P_{a\rightarrow \gamma}\simeq \left(\frac{g_{a\gamma}{\cal{B}}}{m_a}\right)^2 \sim 10^{-35}\left(\frac{{\cal{B}}}{300~ G}\right)^2. 
\label{conversion2}
\ee
The number of the produced photons (as a result of the conversion  from the axions) per unit volume with the frequency  $\omega=m_a$   is estimated as follows
\be
\frac{d N (a\rightarrow \gamma)}{dV} \simeq  n_{\odot}^{\rm axions}  \cdot P_{a\rightarrow \gamma}\simeq   10^{-18}\left(\frac{{\cal{B}}}{300~ G}\right)^2
\left(\frac{\xi}{10^{-10}}\right) \left(\frac{10^{-5} {\rm eV}}{m_a}\right) \frac{1}{\rm cm^3} 
\label{conversion3}
\ee
where $n_{\odot}^{\rm axions}$ is estimated in (\ref{number_density}). These converted photons obviously can   leave the system. The total number of photons leaving the system through  area $\sim 4\pi (2R_{\odot})^2$  per unit time is given by  
\be
\frac{d \Phi (a\rightarrow \gamma)}{ dt}=\frac{d N (a\rightarrow \gamma)}{dV}      \left[4\pi (2R_{\odot})^2\right] c \simeq   10^{16}\left(\frac{{\cal{B}}}{300~ G}\right)^2
\left(\frac{\xi}{10^{-10}}\right) \left(\frac{10^{-5} {\rm eV}}{m_a}\right) \frac{1}{\rm  s}. 
\label{conversion4}
\ee
These photons are very monochromatic with $\omega=m_a$ with accuracy of order $10^{-3}$.  Potentially, it gives us some chance  to observe them on Earth. 
The corresponding count of photons $d F (a\rightarrow \gamma)$ arriving from the Sun with monochromatic frequency  $\omega=m_a$ 
(due to the axion-photon conversion) is estimated as 
\be
  \label{conversion5}
 \frac{d F (a\rightarrow \gamma)}{dA\cdot dt}  \sim \frac{ {d \Phi (a\rightarrow \gamma)}/{ dt}}{4\pi   D^2_{\odot}}\sim  10^{-12}\left(\frac{{\cal{B}}}{300~ G}\right)^2
\left(\frac{\xi}{10^{-10}}\right) \left(\frac{10^{-5} {\rm eV}}{m_a}\right) \frac{1}{\rm cm^2\cdot  s}.
  \ee
 
  This count, of course, is extremely low.  However, these estimates were based on   rate (\ref{conversion2})   corresponding 
   $a\rightarrow \gamma$ conversion in vacuum. As it is known since   \cite{Raffelt:1987im} the rate could be drastically enhanced if the system is placed in a media with non-vanishing plasma frequency $\omega_p$  exactly matching  the axion mass, i.e. $\omega_p=m_a$, which represents the topic for the next subsection.

  \subsection{Solar Corona background. Resonance conversion in solar plasma\label{sect:resonance}} 
  We start with numerical estimation for the plasma frequency $\omega_p$ in the solar corona where the most axions are released as a result of the AQN's annihilation events, 
  \be
  \label{plasma}
  \omega_p\equiv \sqrt{\frac{4\pi\alpha  n}{m_e}}\simeq 3.5\cdot 10^{-6}\cdot\left(\frac{n}{10^{10} {\rm cm}^{-3}}\right)^{\frac{1}{2}} ~~{\rm eV}.
  \ee
  The numerical similarity between $\omega_p$ and the expected  value for the  axion mass $m_a$ from allowed window $m_a\in(10^{-6}-10^{-3})~ {\rm eV}$ represents  the basic motivation for  analysis in this subsection.  In other words,  our goal here is to study possible observational consequences of  the resonance case when the condition $\omega_p= m_a$ could occur in the corona, which is explicit manifestation of the so-called ``level-crossing effect" as formulated in ref.\cite{Raffelt:1987im}.

  If the condition $\omega_p=m_a$ is fulfilled the corresponding resonance $a\rightarrow \gamma$ conversion in media is determined by formula  \cite{Raffelt:1987im}:
  \be
  \label{resonance}
  P_{a\rightarrow \gamma}=\sin^2(\Delta_M L)  , ~~~ \Delta_M=\frac{{\cal{B}}}{2M} \sin \theta, ~~~ M\equiv g_{a\gamma}^{-1}, ~~~ \cos \theta \equiv \hat{\vec{{\cal{B}}}}\cdot \hat{\vec{k}}
  \ee
  where we adopted the notations for $\Delta_M$ from  \cite{Raffelt:1987im} and expressed the fundamental PQ mass scale  $M$ from  \cite{Raffelt:1987im}  in terms of the original definition for $g_{a\gamma}$. Of course we do not expect that this condition can be exactly satisfied in reality in nature. Furthermore, the oscillation length $l_{\rm deg}=\pi/\Delta_M$ is very long, much longer than the size of the system such that 
  $P_{a\rightarrow \gamma}$ never becomes of order one effect. However, our goal here is different, and  
  we present   formula (\ref{resonance}) exclusively for illustrative purposes to illuminate   the role of the distance scale where the conversion occurs. With this purpose we expand the resonance expression (\ref{resonance}) assuming that $\Delta_M L\ll 1$ and compare with non-resonance case (\ref{conversion2}) to arrive
   \be
  \label{resonance1}
  P_{a\rightarrow \gamma}  \simeq (\Delta_M L)^2\simeq  \left(\frac{g_{a\gamma}{\cal{B}}}{m_a}\right)^2\cdot \left(\frac{m_a L}{2}\right)^2,
    \ee
    where we consider special case $\theta=\pi/2$ to simplify the arguments.
Formula (\ref{resonance1}) illustrates very  important point: small conversion rate in non-resonance case (\ref{conversion2})
corresponds to very short distance $ \sim m_a^{-1}$ where this conversion  takes place.  Indeed, the first brackets in (\ref{resonance1}) identically coincides with formula   (\ref{conversion2}) describing the  conversion in non-resonance case. Precisely this first term describes a  huge suppression factor. 

For our present studies     it is important to emphasize that the same formula (\ref{resonance1}) also explicitly shows 
that this suppressed conversion  (\ref{conversion2})  can be greatly enhanced with the second factor $\sim {(m_a L)}^2$ if one can increase the coherence length $L$ by maintaining  $\omega_p=m_a$. If the coherence can be maintained on much larger scale than $m_a^{-1}$ such that  $  {(m_a L)}^2\gg 1 $ the effect of conversion $P_{a\rightarrow \gamma}$ will be strongly enhanced in comparison with  (\ref{conversion2}) by this large factor  $ {(m_a L)}^2\gg 1 $ entering formula 
  (\ref{resonance1}). It is clear  that one should not expect that the effect could be  of order one as one  cannot maintain the coherence on the huge scale $l_{\rm deg}=\pi/\Delta_M$. However, some enhancement in comparison with  (\ref{conversion2}) still can be achieved. 
  
  The same conclusion also follows from the following expression   which was derived using the perturbation theory by treating the inhomogeneities of the  magnetic  field  and plasma density as small perturbations  \cite{Raffelt:1987im}
   \be
  \label{resonance2}
  P_{a\rightarrow \gamma}= \left|\int_0^L dz \Delta_M (z)  \cdot \exp \left(i\Delta_a z-i\int_0^{z} dz' \Delta_{||}(z')\right) \right|^2, ~~~
   \Delta_a=-\frac{m_a^2}{2\omega}, ~~~ \Delta_{||} =-\frac{\omega_p^2}{2\omega}, ~~~~~~
      \ee
  where we neglected $ \Delta_{||}^{\rm vac}\sim B^2$ which is numerically very small for  relatively weak typical  solar magnetic field. 
  
  The conversion rate given by eq. (\ref{resonance2}) generates the enhancement proportional to the large  length $L$ if
  the phases  maintain the coherence and the cancellations between different phases do not occur due to the fast fluctuations. The requirement  that the  coherence is maintained up to the scale  $L$ is determined by the following  condition
  \be
  \label{coherence}
  \left(\Delta_a L-\int_0^{L} dz' \Delta_{||}(z')\right) \lesssim \pi.
  \ee
  If this condition is fulfilled then  the conversion rate given by eq. (\ref{resonance2}) reduces to our previous expression (\ref{resonance1}) with enhancement factor $\sim L^2$, i.e.
   \be
  \label{resonance3}
  P_{a\rightarrow \gamma}  \sim (\Delta_M L)^2 \sim L^2.
    \ee
 This supports our previous conclusion  that   the enhancement factor  $ ({m_a L})^2$ is  a result of constructive  interference.  The corresponding length scale $L$ is determined by condition (\ref{coherence}). 
    
    Now we want to address the  following question: What is the typical length scale $L$ where the condition (\ref{coherence}) can be satisfied in solar atmosphere? We limit our analysis with the trapped axions which have non-relativistic velocities with $\omega\simeq m_a$ as discussed in previous section \ref{non-resonance}. These axions are distributed in the entire solar atmosphere. Therefore, there is always an extended  region in corona or chromosphere where the electron density $n$ is such 
    that the plasma frequency (\ref{plasma}) equals the axion mass, i.e. $m_a=\omega_p$. The only question remains to be answered is what are the typical length scales where the average value $\la n \ra$ for the electron density varies\footnote{Local fluctuations of the density always occur as a result of different types of waves, including the sound waves,  in plasma. However, it is expected that these oscillations do not change the integral entering (\ref{coherence}). In other words, we are interested in steady and sustained  variation of the average density $\la n \ra$ with latitude and altitude, rather than numerous conventional   fluctuations which always occur in hot plasma but do not modify the average magnitude of the integral   (\ref{coherence}).}.

    To estimate the corresponding scale $L$ we notice that a typical variation of the   density (and the plasma frequency) 
    $n$ by factor $\sim 10$ occurs when the  altitude changes by  $\sim 10^3$ km. Assuming a linear extrapolation (excluding very fast changes in the transition region) one should expect that the variation of the density $\delta n/n \sim1 $  occurs on the scale of order $l_0\sim 10^2$ km. This estimate implies that the relative variation  (mismatch) of the density
    on the coherence scale $L$ of the axion/photon oscillation must not exceed $\lambda/L$ to be consistent  with (\ref{coherence}).
    In other words, the scale $L$ where coherence  (\ref{coherence}) can be maintained must satisfy the following condition
    \be
    \label{L}
    L\sim l_0\frac{\lambda}{L} ~~~ \Rightarrow ~~~  L\sim \sqrt{\lambda l_0}\sim 0.4\cdot 10^4  \sqrt{ \frac{10^{-5} {\rm eV}}{m_a}}~{\rm cm}.
    \ee
    Precisely at this coherence scale $L$ the mismatch in plasma frequency  is sufficiently small as $(\delta n/n)_L\sim  \lambda/L$. At the same time     $(\delta n/n)_{ l_0}\sim  (\lambda/L)\cdot (l_0/L)\sim 1$ becomes order of one at much larger scales $\sim l_0$ where coherence, of course, cannot be maintained. One should emphasize that this very rough estimate assumes a linear extrapolation. This assumption may or may not be justified in reality in  solar atmosphere. One should emphasize here that any variation of the magnetic field entering (\ref{resonance2}) do not modify our estimate for the coherence length (\ref{L}). This is  because the estimate (\ref{L}) is sensitive to the phase variation (rather than to  the amplitude  changes $\sim\Delta_M(z)$) determined by a steady and systematic variation of the plasma frequency $\omega_p\sim \sqrt{n}$ in corona.

    If one literally  accepts the estimate  (\ref{L})    the  corresponding   enhancement factor  can be approximated   as follows,
    \be
    \label{enhancement}
    (Lm_a)^2\sim 4\cdot 10^6  \left(\frac{m_a}{10^{-5} {\rm eV}}\right) \gg 1. 
    \ee
    Our previous (non-resonance) estimate (\ref{conversion5}) should be multiplied by the enhancement factor  (\ref{enhancement})
    for case if the resonance conditions can be satisfied and the linear extrapolation is justified. The rate of conversion with this factor becomes
     \be
  \label{conversion6}
 \frac{d F (a\rightarrow \gamma)}{dA\cdot dt}\Big |_{\rm resonance}  \approx  10^{-6}\left(\frac{{\cal{B}}}{300~ G}\right)^2
\left(\frac{\xi}{10^{-10}}\right)   \frac{1}{\rm cm^2\cdot  s}.
  \ee
  The  corresponding energy flux  can be estimated as 
       \be
  \label{conversion7}
 m_a\frac{d F (a\rightarrow \gamma)}{dA\cdot dt}\Big |_{\rm resonance}  \approx  10^{-30}\left(\frac{{\cal{B}}}{300~ G}\right)^2
\left(\frac{\xi}{10^{-10}}\right) \left(\frac{m_a}{10^{-5} {\rm eV}}\right)   \frac{ W}{\rm cm^2},
  \ee
  where we expressed the intensity in conventional  $(W/{\rm cm}^2)$ units using the relations $1~W= 10^7({\rm erg/s})$ and $1~{\rm eV}=1.6\cdot 10^{-12} {\rm erg}$.
  
While the count (\ref{conversion6}), (\ref{conversion7}) is still very low, 
  some  hope is that this is an unique monochromatic line. Furthermore, the intensity of this line must be correlated with the EUV emission from corona. In addition, during the flares the magnetic field ${\cal{B}}$ might be very large    in the solar system 
  which provides some enhancement factor  and possible correlations with the flares. Finally, this monochromatic line can be, in principle, discriminated from the background noise 
  as it should appear only along the line-of-sight in the direction of the Sun.
  
  It is very instructive to compare the intensity (\ref{conversion7}) with corresponding conventional energy flux from the Sun in the frequency band $\omega\approx m_a$. To proceed with the estimates we  recall that the total solar intensity (integrated over all frequency bands)  at Earth surface is about $0.14\cdot  \frac{ W}{\rm cm^2}$, which of course many orders of magnitude higher than the rate (\ref{conversion7}). However, we should compare (\ref{conversion7}) not with the total intensity from the Sun measured on Earth, but rather with the solar energy flux from the low energy frequency band 
  with $\omega\leq m_a$. The corresponding estimates can be easily performed as the corresponding spectral properties are determined by the Reyleigh- Jeans formula for low energy tail of the black-body (BB) radiation, i.e.
  \be
  \label{BB}
  dE_{\omega}=\frac{T}{\pi^2}\omega^2 d\omega, ~~~ E^{\rm BB}_{\rm tot}=\frac{\pi^2}{15}T^4, ~~~ \frac{E_{\omega\leq m_a}}{E^{\rm BB}_{\rm tot}} \sim \frac{5}{\pi^4}\left(\frac{m_a}{T}\right)^3.
  \ee
  The intensity of the conventional BB  solar radiation in the low energy frequency band with $\omega\leq m_a$
  is estimated as follows
  \be
  \label{BB1}
\left( 0.14\cdot  \frac{ W}{\rm cm^2}\right)\cdot  \frac{5}{\pi^4}\left(\frac{m_a}{T}\right)^3\sim 0.4 \cdot 10^{-16}  \left(\frac{m_a}{10^{-5} {\rm eV}}\right)^3 \frac{ W}{\rm cm^2}, 
  \ee
  which is still much higher  than the   energy flux due to the axion conversion (\ref{conversion7}) within conventional axion mass window. Only for ultra light axions with $m_a\leq 10^{-12} ~{\rm eV}$ the background (\ref{BB1}) becomes below the signal (\ref{conversion7}). The corresponding case of ultra light axion is not part of this work and shall  not be further elaborated. 
  
  Therefore, one should completely remove   the emission from the photosphere with  large background (\ref{BB1}) for analyzing  of the axion conversion (\ref{conversion7}) from corona. Such removing   occurs naturally during the solar eclipses. In practice, astronomers  in the past have developed a number of  technical tools which  allowed to study  a weak emission from corona by  removing a much stronger   radiation  from photosphere. 
   A  high resolution instrument  would be  very beneficial  to study  the monochromatic   line (\ref{conversion7}) with the width $\Delta \omega$, which   is determined by the escape velocity  $ v^{\rm trapped}_{\odot} $ from (\ref{trapped}), i.e.
 $\Delta \omega\sim m_a v^{\rm trapped}_{\odot} \sim 10^{-3} m_a$. 
     
     To recapitulate: the expected signal from the solar corona  is very low as our estimate (\ref{conversion7}) suggests. 
     It can be only  studied if the background  radiation (\ref{BB1}) from photosphere can be removed from analysis and  the instrumental resolution is sufficiently  high to  study   the highly monochromatic emission (with the width $\Delta \omega/\omega\sim 10^{-3}$)  from corona.
          Needless to say that a strong magnetic field in a detector  is not required  for the observation of these photons on Earth because the axion-photon conversion occurs  in the solar atmosphere rather than on Earth. In a sense we use entire Sun as a one big helioscope where the trapped axions have been accumulated during 4.5 billion years and where the axions can be  converted  to photons in the entire solar atmosphere.
 
 In next subsection we consider much more optimistic 
     case when the axions are trapped by the Earth. As we shall see below, the density of such axions  could be the same order of magnitude as the  galactic axion density, which is the conventional normalization point for the most  presently operational  (or under construction, or  in stage  of design) axion search experiments.

 
  \subsection{Axions from the Earth's  Underground}\label{earth_under}
  According to ref.  \cite{Fischer:2018niu} the axion flux due to the AQN annihilation events  in the very deep underground 
  is given by (\ref{earth-axion}).  We integrate this rate over entire surface to arrive   
   \be
  \label{axions_trapped_1}
  \frac{d E_{\oplus}}{dt} ({\rm trapped ~axions})    \sim  10^{16}\xi_{\oplus}\cdot \left(\frac{\Delta B}{B}\right)\cdot 4\pi R^2_{\oplus}\frac{\rm eV}{\rm s} \sim 10^{18} \cdot  \left(\frac{\xi_{\oplus}}{10^{-17}}\right) \cdot  \left(\frac{\Delta B}{B}\right)  \frac{\rm eV}{\rm  s}
  \ee
  where we  used the expression  (\ref{axions_rate}) for the rate of the energy transfer to the axions. We also     introduced the suppression factor $\xi_{\oplus}$ to account for the small fraction of the trapped axions with $v\leq {v_{\rm trapped}}$. For numerical estimates  in formula  (\ref{axions_trapped})   we use   suppression factor $\xi_{\oplus}\sim \xi_{\odot}\cdot (v_{\oplus} /v_{\odot})^4\sim 10^{-17}$ computed in previous section and given by (\ref{trapped}). 
  This is of course very tiny rate even when  $\Delta B/B\sim 1$ as we expect.

    The axions  (\ref{axions_trapped_1}) could not  leave the system  during   entire life time of the Earth, i.e. 4.5 billion years $\simeq 10^{17}{\rm s}$. Therefore, the total energy accumulated  by the Earth   and related to AQN annihilation events   radiating  the slow velocity  axions can be estimated as follows
  \be
  \label{axions_total_1}
   E_{\oplus}  ({\rm trapped ~axions})  \simeq 10^{18} \cdot  \left(\frac{\xi_{\oplus}}{10^{-17}}\right)  \left(\frac{\Delta B}{B}\right) \cdot  \frac{\rm eV}{\rm  s}\cdot 10^{17}s\simeq 10^{35}  \left(\frac{\xi_{\oplus}}{10^{-17}}\right) \left(\frac{\Delta B}{B}\right) {\rm eV}.
  \ee
  This energy can be expressed  in terms of extra Earth's mass $\Delta M_{\oplus}$  accumulated  by the Earth and represented by the trapped axions
   \be
  \label{axions_mass_1}
 \Delta M_{\oplus}   ({\rm trapped ~axions})  \simeq  0.1 \left(\frac{\xi_{\oplus}}{10^{-17}}\right) \left(\frac{\Delta B}{B}\right) {\rm kg},
  \ee 
  which of course represents a very tiny fraction of the Earth mass $ M_{\oplus}\simeq 5.9\cdot 10^{24} {\rm kg}$.
  
The energy (\ref{axions_total_1}) corresponds to the following total number of the axions accumulated  by the Earth during  its life-time:
 \be
  \label{number_total_1}
   N_{\oplus}^{\rm axions} \sim \frac{E_{\oplus}  ({\rm trapped ~axions}) }{m_ac^2} \simeq 10^{40} \cdot   \left(\frac{\xi_{\oplus}}{10^{-17}}\right) \left(\frac{\Delta B}{B}\right) \cdot \left(\frac{10^{-5} {\rm eV}}{m_a}\right).
  \ee
If we assume that the majority of these axions are localized within    radius $R_{\oplus}$, we arrive to the following estimate for the average axion energy density inside this volume
\be
  \label{energy_density_1}
 \rho_{\oplus}^{\rm axions}\sim \frac{E_{\oplus}  ({\rm trapped ~axions})}{\frac{4}{3}\pi R_{\oplus}^3} \sim   0.1      \left(\frac{\xi_{\oplus}}{10^{-17}}\right)  \left(\frac{\Delta B}{B}\right)  \frac{\rm GeV}{\rm cm^3}.
  \ee
which is amazingly close  to  the  average dark matter density today $\rho_{\rm DM}\simeq 0.3 ~ \frac{\rm GeV}{\rm cm^3} $.
The eq. (\ref{energy_density_1}) should be viewed as the order of magnitude estimate at the very best. The main uncertainty 
here is that the trapped axions are not distributed uniformly, as assumed in (\ref{energy_density_1}). Instead, they are obviously distributed  in a highly 
 nontrivial way determined by the position of the nugget when emission occurs (in deep underground) and the direction of the velocity at the moment of emission. 
 Though the estimate  (\ref{energy_density_1}) is rough, it is also very promising as it suggests that the galactic  axion density and the axion density produced by the AQN mechanism could be the same order of magnitude. 
 \exclude{In addition, the contribution of the galactic axions to the dark matter density scales as $m_a^{-7/6}$ as mentioned in Introduction.
 It should be contrasted with estimate (\ref{energy_density_1}) which is not sensitive to the value of the axion mass $m_a$ as a result of very generic feature expressed as  $\Omega_{\rm dark}\sim \Omega_{\rm visible}$ in  the AQN framework.  
}  
  
  One may wonder\footnote{We are thankful to anonymous Referee who suggested to produce such estimates.}  if the  terrestrial geomagnetic   field can be used as the axion converter, and if the resonance conversion may occur on Earth, similar to our discussions in previous Sect. \ref{sect:resonance}
  devoted to the resonance conversion in solar corona. Unfortunately, for the conventional     axion mass window $m_a\in(10^{-6}-10^{-3})~ {\rm eV}$ the resonance conditions cannot be satisfied. Indeed, while the Earth's ionosphere is highly ionized, 
  the corresponding electron density is very low: $n\sim 10^6 {\rm cm^{-3}}$ for the so-called F-layer which extends from the altitude 150 km for few hundred kilometers. The corresponding plasma frequency in ionosphere 
   \be
  \label{plasma1}
  \omega_p\equiv \sqrt{\frac{4\pi\alpha  n}{m_e}}\simeq 3.5\cdot 10^{-8}\cdot\left(\frac{n}{10^{6} {\rm cm}^{-3}}\right)^{\frac{1}{2}} ~~{\rm eV}
  \ee
is well below the typical axion mass. This estimate suggests that the resonance case  cannot be realized  for the conventional  axion mass window.  Therefore,   
one should use non-resonance  formula for conversion:
  \begin{equation}
\label{eq:4. earth_monochromatic}
m_a\frac{d F_\oplus (a\rightarrow \gamma)}{dA\cdot dt}
\simeq\frac{1}{2}\rho_\oplus^{\rm axions} P_{a\rightarrow\gamma}c
\simeq 10^{-41}\left(\frac{\xi_{\oplus}}{10^{-17}}\right)
\left(\frac{\Delta B}{B}\right)\left(\frac{{\cal{B}}}{0.5~ G}\right)^2 
\rm\frac{W}{cm^2}.
\end{equation}
This estimate indicates that the corresponding rate is  too low to be observed if one uses the terrestrial geomagnetic   field as the axion converter. The crucial suppression factors here are the small terrestrial geomagnetic magnetic field ${\cal{B}}\sim 0.5$ G and very small escape velocity (\ref{trapped}) which leads to the very tiny portion of the trapped axions $\xi_{\oplus}\sim 10^{-17}$. 
Similar arguments also apply to other planets such as  Jupiter, which  has no resonant enhancement nor sufficiently strong magnetic field comparable to the solar sunspots. Therefore, while the AQNs obviously get annihilated in the Jupiter's underground 
producing additional internal heat, the corresponding axion emission would be 
even  smaller than from the Sun  (\ref{conversion7}) due to a number of additional suppression factors such as  smaller mass (and therefore, smaller impact parameter leading to a smaller AQN flux hitting Jupiter), larger distance from Earth,  smaller escape velocity (leading to a smaller parameter $\xi$) in comparison with the solar  $\xi_{\odot}$, etc. 

   Therefore we return to our main and most promising estimate  (\ref{energy_density_1}) which indicates that
   the  density of the bound axions could be the same order of magnitude as the galactic DM axions, and therefore 
   the conventional instruments originally designed for the galactic axion searches  can be also used to study the trapped axions. 
   The distinct feature of the AQN trapped axions is very large wave length $\lambda_a=(m_av_a)^{-1}$ as the typical trapped velocity $v_a\leq v^{\rm trapped}_{\oplus}$ is much smaller  than a typical galactic DM velocity $\sim 10^{-3}c$ according to (\ref{trapped}). This unique feature of the trapped axions might be the ``smoking gun" leading to their discovery.

\section{Conclusion and future directions\label{conclusion}}
This work represents a natural generalization of the previous studies \cite{Fischer:2018niu} to 
properly account for the production of  the low energy axions when the AQNs get annihilated in the Sun or Earth
and emit axions with $v\leq {v^{\rm trapped}_{\odot}}$ in the  solar corona or $v\leq {v^{\rm trapped}_{\oplus}}$
in the deep Earth's underground. This portion of the non-relativistic axions is extremely tiny as shown on Fig. \ref{fig:4.2 3D_rhovE2_del00}.  However,  the effect is  drastically enhanced  as argued  in Section \ref{lensing} due to  accumulation of these axions during entire life of the solar system, i.e. for $\sim $4.5 billion of years. The corresponding estimates represent the main results of the present studies.

We shall not repeat and discuss here  a large number of  estimates     presented  in  Section \ref{lensing}. Instead, we focus on a single formula  
(\ref{energy_density_1})  describing  the  energy  density of the trapped axions  $\rho_{\oplus}^{\rm axions}$. We think this estimate  has   a   huge discovery potential  because $\rho_{\oplus}^{\rm axions}$ is relatively large and comparable with the  average galactic dark matter density today $\rho_{\rm DM}\simeq 0.3 ~ \frac{\rm GeV}{\rm cm^3} $. What is more important is that the spectral features of the trapped axions are very distinct from conventional galactic axions because the typical trapped velocity $v_a\leq v^{\rm trapped}_{\oplus}$ is much smaller  than a typical galactic DM velocity $\sim 10^{-3}c$ according to (\ref{trapped}). Therefore, the typical       wave length $\lambda_a=(m_av_a)^{-1}$ of these axions is much longer  in comparison with galactic axions. This unique feature makes the trapped axions are very distinct from conventional galactic axions.     These axions obviously can be easily discriminated from anything else. The discovery of such axions would be a ``smoking gun" for the entire AQN proposal unifying
the DM  and baryogenesis (separation of charges) problems.      

 This  new mechanism of the axion production is entirely based on the unorthodox AQN dark matter  model. 
  Why we think that this new AQN framework (and  accompanying  the axion emission)  should be taken seriously? We refer to  \cite{Fischer:2018niu}  for overview of this DM model. Nevertheless, we want to make  few  comments  
   here  suggesting that the AQN framework should be indeed taken seriously.
   
   We start with the remark  that this model was invented long ago as a natural 
    explanation of the observed ratio   $\Omega_{\rm dark}\sim \Omega_{\rm visible}$  between visible and dark matter 
densities.  In context of the present work the most important 
    feature of this model is that it   may potentially resolve the old renowned puzzle (since 1939) known in the community under the name ``the Solar Corona Mystery".  In particular, this model, without adjusting any parameters, generates the observed EUV luminosity (\ref{estimate}) as reviewed in \cite{Fischer:2018niu}. 
       
       Furthermore,  the AQN resolution of the solar corona puzzle also resolves another mystery  \cite{Zioutas} where 
 it was  claimed   that a number of highly unusual phenomena   observed in solar atmosphere might be related  to the gravitational lensing of ``invisible" streaming matter towards the Sun which is correlated with positions of the planets. This is really a  weird correlation because  one should not  expect    any  connections  between the   flare occurrences, the intensity of the EUV radiation,    and the position of the planets. 
 At the same time,
  such ``weird" correlations    naturally occur  within AQN framework. This is because the dark matter AQNs, being the ``invisible streaming matter" (in terminology of ref.   \cite{Zioutas}) can play the role of the triggers sparking the large flares \cite{Zhitnitsky:2018mav}. Therefore, the observation of the correlation between the EUV intensity, the  frequency of the flares  and positions of the planets can be considered as an additional supporting argument  of  the dark matter explanation of the observed EUV irradiation  (\ref{estimate}), 
because both effects are originated from the same dark matter AQNs.

Last, but not least.  The AQN  model offers a very natural resolution of the so-called ``Primordial Lithium Puzzle" as recently argued in 
\cite{Flambaum:2018ohm}. This problem  has been with us for at least two decades, and   conventional astrophysical and nuclear physics  proposals could not resolve this longstanding mystery. In the AQN framework this puzzle is automatically and naturally resolved without adjusting any parameters as shown in \cite{Flambaum:2018ohm}. This resolution represents yet another, though indirect, support for this new AQN framework. 

All these  arguments obviously represent {\it indirect} support for the AQN paradigm. The discovery  of the trapped axions with energy density (\ref{energy_density_1}) and with drastically distinct spectral features  (in comparison with conventional galactic axions) would be the {\it direct} support for this model as it is hard to imagine any other model which could produce the axions with  $v_a\leq v^{\rm trapped}_{\oplus}$ with sharp cutoff in density for $v_a>v^{\rm trapped}_{\oplus}$. We conclude this work  on this optimistic note.

\section*{Acknowledgements} 
One of us (AZ) is thankful to Konstantin Zioutas for enormous  number of questions which   motivated these studies. 
We are thankful to Ludo van Waerbeke for discussions on feasibility and perspective to drastically improve the order of magnitude estimates (\ref{energy_density}), (\ref{energy_density_1}) using   numerical Monte Carlo simulations accounting for all possible axion orbits. 
 This research was supported in part by the Natural Sciences and Engineering 
Research Council of Canada.

\appendix
\section{Technical details. Axion emission from the domain wall. 3D case}
\label{Appendix_3D}

In this Appendix we want to study the spectral properties of the axion's emission as a result of time-dependent perturbations of the axion domain wall.
We want to focus on the axion portion of the axion DW, which also includes other fields such as $\pi, \eta'$,  see \cite{Forbes:2000et}. It   also contains a phase describing the baryon charge distribution on the surface of the nugget as discussed in 
\cite{Liang:2016tqc}. Exact features of the profile functions for all these fields  are not important for our purposes. Therefore, one can simplify our computations by considering the  following  
effective Lagrangian with two  degenerate vacuum states\footnote{\label{footnote:A qualitative model}In our previous studies  \cite{Liang:2016tqc,Ge:2017ttc,Ge:2017idw} we always discussed the so-called $N=1$ domain walls. It implies that   the vacuum is unique and the DW solution interpolates between one and the same physical vacuum. This interpolation always occurs as a result of variation of the axion field together  with another fields, such as $\pi$ or $ \eta'$ as discussed in \cite{Forbes:2000et}. These additional fields do not generate much changes in  the domain wall tension, nor they 
affect our analysis of the axion production, which is the subject of the present work. Therefore, we ignore these fields to simplify notations and qualitative analysis in this work.}.  
\begin{equation}
\label{eq:A L(phi)}
{\cal S}[\phi]
=\int d^4x\left[\frac{1}{2}(\partial_\mu\phi)^2
-\frac{g^2}{4}\left(\phi^2-\frac{\pi^2}{4}f_a^2\right)^2\right],
\end{equation}
where $g=\frac{\sqrt{2}}{\pi}\frac{m_a}{f_a}$, and we set the effective axion angle as $\phi/f_a\equiv\theta+{\rm arg~det}M+\pi/2$ (note that we shift the angle by $\pi/2$ for convenience of calculation). In this work, we are especially interested in the non-relativistic domain where thin-wall approximation is badly violated. Thus, we should approach the solution in 3D case. Since the ground-state domain wall solution must preserve spherical symmetry, the equation of motion reads
\begin{equation}
\label{eq:A equation of motion}
\frac{\partial^2}{\partial r^2}\phi(r)
+\frac{2}{r}\frac{\partial}{\partial r}\phi(r)
=g^2\phi(r)\cdot\left[\phi^2(r)-\frac{\pi^2}{4}f_a^2\right],
\qquad \phi(R_0)=0
\end{equation}
where $R_0$ defines the boundary which separates two distinct physical vacua and it coincides with the radius of the AQN in equilibrium. While the exact domain wall solution to Eq. \eqref{eq:A equation of motion} is hard to solve, the approximate solution gives\footnote{
Note that interaction between axion and other fields such as $\pi$ and $\eta'$ becomes strong within $r\lesssim R_0$, see \cite{Forbes:2000et}. Hence, we should set a cutoff range at $r\lesssim R_0$ where Eq. \eqref{eq:A equation of motion} is no longer valid. 
}
\begin{equation}
\label{eq:A approximate DW solution}
\phi_{w,R_0}(r)\simeq\left\{
\begin{aligned}
&\frac{\pi}{2}f_a
\cdot\frac{R_{\rm eff}}{r}\tanh\left[\frac{1}{2}m_a(r-R_0)\right],
&R_0\lesssim r\leq R_{\rm trans}  \\
&\frac{\pi}{2}f_a\cdot\tanh\left[\frac{1}{2}m_a(r-R_0+\delta_R)\right],
&r>R_{\rm trans}
\end{aligned}
\right.
\end{equation}
where $R_{\rm eff}$ and $\delta_R$ are functions of a tunable parameter $R_{\rm trans}$
\begin{subequations}
\label{eq:A approximate DW solution_ass delta_R and R_eff}
\begin{equation}
\label{eq:A approximate DW solution_ass delta_R}
\delta_R
\simeq\frac{1}{R_0}(R_{\rm trans}-R_0)^2
\equiv\frac{1}{m_a}\delta,
\end{equation}
\begin{equation}
\label{eq:A approximate DW solution_ass R_eff}
R_{\rm eff}
\simeq R_{\rm trans}\frac{\tanh[
\frac{1}{2R_0}m_aR_{\rm trans}^2]}{\tanh[\frac{1}{2}m_a R_{\rm trans}]}
\end{equation}
\end{subequations}
within domain $R_0<R_{\rm trans}\lesssim\sqrt{m_aR_0}\cdot m_a^{-1}$. 
One can explicitly check this approximate solution \eqref{eq:A approximate DW solution} is continuous and first order differentiable. Also, it is precisely the exact solution in the near-field limit $r\sim R_0$ and the far-field limit $r\gg m_a^{-1}$. Hence, the only unknown part the solution is the ``transition'' regime between these two limits, where we introduce a tunable parameter $R_{\rm trans}$ to account for this type of error source. We will keep this parameter in the following calculations, so it serves as a probe to test whether the final result is sensitive to our crude approach in the transition regime. As we will see, the final result is   not sensitive to the tuning of $R_{\rm trans}$. 

Lastly, instead of using $R_{\rm trans}$ directly, it is more convenient to define a simple parameter $\delta\equiv m_a\delta_R$ which roughly varies from 0 to 1. As we will see, $\delta$ is the only parameter entering the final result. 

We are now ready to compute the excitations $\chi(t,z)$ in the time dependent background. These excitations will be eventually identified as the axions emitted by the axions DW. To achieve this task we expand $\phi(t,z)=\phi_w(z-R_0)+\chi(t,z)$, which gives 
\begin{equation}
\label{eq:A action_chi excitation}
{\cal S}[\phi]
={\cal S}[\phi_w]
+\int dt\int d^3x \left[
\frac{1}{2}\dot{\chi}^2-\frac{1}{2}\chi L_2\chi
\right]+{\cal O}(\chi^3).
\end{equation}
where $L_2$ is a linear differential operator of the second order,
\begin{equation}
\label{eq:A action_chi excitation_ass L2}
\begin{aligned}
L_2\chi
&=\left.
-\frac{1}{r}\frac{\partial^2(r\chi)}{\partial r^2}
-\frac{1}{r^2}\left[
\frac{1}{\sin\theta}\frac{\partial}{\partial\theta}
\left(\sin\theta~\frac{\partial \chi}{\partial\theta}\right)
+\frac{1}{\sin^2\theta}\frac{\partial^2\chi}{\partial \phi^2}
\right]
+\left[2g^2\phi^2\chi+g^2(\phi^2-v^2)\chi
\right]\right|_{\phi=\phi_{w,R_0}}  \\
&=-\frac{1}{r}\frac{\partial^2(r\chi)}{\partial r^2}
-\frac{1}{r^2}\left[
\frac{1}{\sin\theta}\frac{\partial}{\partial\theta}
\left(\sin\theta~\frac{\partial \chi}{\partial\theta}\right)
+\frac{1}{\sin^2\theta}\frac{\partial^2\chi}{\partial \phi^2}
\right]
+\frac{1}{2}\frac{m_a^2}{v^2}(3\phi_{w,R_0}^2-v^2)\chi.
\end{aligned}
\end{equation}
The corresponding equation of motion is therefore
\begin{equation}
\label{eq:A chi PDE}
\frac{\partial^2}{\partial t^2}\chi=-L_2\chi.
\end{equation}
To look for the initial conditions, we now want to describe the emission of axions in one cycle of oscillation. As mentioned in Sec. \ref{sec:4.2 Spectral properties}, annihilation of baryon charge results in oscillations of domain wall. Assuming the oscillation is approximately adiabatic, it is sufficient to only analyze the first half of an oscillation -- say, the ``contraction period''-- where the domain wall shrinks from $R_0$ to a slightly smaller size $R_0-\Delta R$. We assumed the rest half of the cycle, the ``expansion period'', is just the time-reversed and produces an equivalent contribution. We may write down such initial conditions as
\begin{subequations}
\label{eq:A phi_IC}
\begin{equation}
\label{eq:A phi_IC1}
\phi(0,r)=\phi_{w,R_0}(r)
\end{equation}
\begin{equation}
\label{eq:A phi_IC2}
\begin{aligned}
&\phi(\frac{1}{2}t_{\rm osc},r)=\phi_{w,R_0-\Delta R}(r)+({\rm excitations})
\end{aligned}
\end{equation}
\end{subequations}
where $t_{\rm osc}$ denotes the period of one full oscillation. The excitation modes in condition \eqref{eq:A phi_IC2} is unknown and depends on the conversion rate from excitation modes to freely propagating axions. In terms of $\chi$, the initial conditions \eqref{eq:A phi_IC} imply
\begin{subequations}
\label{eq:A chi_IC}
\begin{equation}
\label{eq:A chi_IC1}
\chi(0,r)=0
\end{equation}
\begin{equation}
\label{eq:A chi_IC2}
\begin{aligned}
\chi(0,r)
=\eta(\theta,\varphi)\partial_{R_0}[\phi_{w,R_0}(r)]\Delta R
+{\cal O}(\Delta R^2)
\end{aligned}
\end{equation}
\end{subequations}
where we introduce a free parameter $\eta(\theta,\varphi)$ which may be interpreted as the ``amplitude of efficiency'' of the conversion rate from excitations to free axions, so $\eta$ must vary between 0 to 1. However, $\eta$ here may be also interpreted as a correction term like $\delta$ in the approximate solution  \eqref{eq:A approximate DW solution} within the transition regime $R_0\ll r\lesssim m_a^{-1}$, so $\eta$ can be greater than 1 in general. Nonetheless, we will expect $\eta\sim1$ and will treat it as a normalization factor regarding to the luminosity. And in general, $\eta$ can be expanded by partial waves
\begin{equation}
\label{eq:A IC_2_ass eta}
\begin{aligned}
&\eta(\theta,\varphi)
=\sum_{l=0}^{\infty}\sum_{m=-l}^{l}\eta_{lm}Y_{lm}(\theta,\varphi),  \\
&\eta_{lm}
=\int_0^{2\pi}d\varphi\int_0^\pi d\theta\sin\theta~
Y_{lm}^*(\theta,\varphi)\eta(\theta,\varphi).
\end{aligned}
\end{equation}
If we assume a good spherical symmetry preserves during the most period of the annihilation process of AQN, then $\eta_{00}$ will be the dominant contribution and $\eta_{10}$ be the next order correction. 

To solve for the excitation mode, it is convenient to write $\chi$ in terms of some normalized basis. The expansion for free wave is conventionally
\begin{equation}
\label{eq:A chi in chi_plm_main}
\chi(t,r,\theta,\varphi)
=\sum_{l=0}^{\infty}\sum_{m=-l}^{l}\int d^3p~
a_{plm}(t)\chi_{plm}(r,\theta,\varphi),\quad
\chi_{plm}(r,\theta,\varphi)
=\frac{1}{\sqrt{4\pi^2E_a}}j_l(pr)Y_{lm}(\theta,\varphi)
\end{equation}
where $j_l(x)$ is the spherical Bessel function, and we have implicitly used two orthogonalities
\begin{subequations}
\label{eq:A orthogonality_JY}
\begin{equation}
\label{eq:A orthogonality_J}
\int_0^\infty dr~r^2j_l(pr)j_l(qr)=\frac{\pi}{2p^2}\delta(p-q),
\end{equation}
\begin{equation}
\label{eq:A orthogonality_Y}
\int_0^{2\pi}d\varphi\int_0^\pi d\theta\sin\theta~
Y^*_{lm}(\theta,\varphi)Y_{l'm'}(\theta,\varphi)
=\delta_{ll'}\delta_{mm'}.
\end{equation}
\end{subequations}
Note that $L_2$ is diagonal in basis of $\chi_{plm}$
\begin{equation}
\label{eq:A L2 matrix components}
\begin{aligned}
\int d^3x \chi_{ql'm'}^*(r,\theta,\varphi)L_2\chi_{plm}(r,\theta,\varphi)
&=\frac{1}{8\pi E_a}\delta(p-q)
+\frac{m_a^2}{4\pi^2E_a}K_{pq}^{(l)}
\int dr~r^2j_l(qr)j_l(pr) \\
&=\frac{\delta(p-q)}{8\pi E_a p^2}(p^2+K_{p,q}^{(l)}m_a^2),
\end{aligned}
\end{equation}
where $K_{p,q}^{(l)}$ is a coefficient defined as
\begin{equation}
\label{eq:A L2 matrix components_ass K}
\begin{aligned}
K_{p,q}^{(l)}
&\equiv \lim_{L\rightarrow\infty}\frac{\int_0^L dr~r^2 j_l(pr)j_l(qr)
\frac{1}{2}\left[
3\left(\dfrac{1}{v}\phi_{w,R_0}(r)\right)^2-1\right]}
{\int_0^L dr~r^2 j_l(pr)j_l(qr)}
\end{aligned}
\end{equation}
for simplicity of calculation. In   \ref{appendix:B About K} we can show $K_{p,q}^{(l)}\delta(p-q)=\delta(p-q)$. Then Eq. \eqref{eq:A chi PDE} is simplified to
\begin{equation}
\label{eq:A L2 matrix components_2}
\frac{d^2}{dt^2}a_{plm}(t)=-E_a^2(p)a_{plm}(t),\qquad
E_a(p)\equiv\sqrt{p^2+m_a^2},
\end{equation}
which clearly has solution
\begin{equation}
\label{eq:A a solution_1}
a_{plm}(t)=b_{plm}\sin E_at,
\end{equation}
following the initial condition \eqref{eq:A chi_IC1}, where $b_{plm}$ is an time-independent coefficient to be determined. To find $b_{plm}$, we should impose the second initial condition \eqref{eq:A chi_IC2} which implies
\begin{equation}
\label{eq:A b_solution_1}
\begin{aligned}
b_{plm}
&=\frac{\pi}{2}\frac{f_am_a\Delta R~\eta_{lm}}{\sin(\frac{1}{2}E_at_{\rm osc})}
\sqrt{\frac{E_a}{4\pi^2}}\left\{
\int_0^{R_{\rm trans}}dr~r^2\cdot
\frac{R_{\rm eff}}{r}{\rm sech}^2\left[
\frac{m_a}{2}(r-R_0)\right]j_l(pr)
\right.  \\
&\left.\qquad\qquad\qquad\qquad\qquad-
\int_0^{R_{\rm trans}}dr~r^2\cdot
{\rm sech}^2\left[\frac{m_a}{2}(r-R_0+\delta_R)\right]j_l(pr)
\right.  \\
&\left.\qquad\qquad\qquad\qquad\qquad+
\int_0^{\infty}dr~r^2\cdot
{\rm sech}^2\left[\frac{m_a}{2}(r-R_0+\delta_R)\right]j_l(pr)
\right\}.  \\
\end{aligned}
\end{equation}
Note that only the last term in the curly bracket is dominant because $R_{\rm trans}\ll m_a^{-1}$ largely suppresses the first two terms.\footnote{\label{footnote:A R hierachy}
More specifically, due to the fact $R_0\ll m_a^{-1}$, we have the hierarchy $R_0<R_{\rm trans}\lesssim\frac{1}{2}\sqrt{m_aR_0}\cdot m_a^{-1}\ll m_a^{-1}$. 
}  Thus, we conclude
\begin{equation}
\label{eq:A b_solution_2}
\begin{aligned}
b_{plm}
\simeq\frac{\pi}{2}
\frac{f_am_a\Delta R~\eta_{lm}}{\sin(\frac{1}{2}E_at_{\rm osc})}
\sqrt{\frac{E_a}{4\pi^2}}\left\{
\int_0^{\infty}dr
~r^2\cdot
{\rm sech}^2\left[\frac{m_a}{2}(r+\delta_R)\right]j_l(pr)
+{\cal O}(R_{\rm trans}^{l+2})
\right\},
\end{aligned}
\end{equation}
where we have also drop $R_0$ in the hyperbolic secant function because it is of order $R_{\rm trans}$. This integral can be evaluated precisely if we expand the hyperbolic secant as
\begin{equation}
\label{eq:A cluster expand sech}
\begin{aligned}
{\rm sech}^2(\frac{1}{2}x)
&=e^{-x}\sum_{n=0}^{\infty}(-1)^n(n+1)\frac{1}{2^n}(e^{-x}-1)^n  \\
&=\sum_{n=0}^{\infty}\sum_{k=0}^{n}
\frac{1}{2^n}\frac{(n+1)!}{k!(n-k)!}(-1)^ke^{-(k+1)x}
\end{aligned}
\end{equation}
and use the fact
\begin{equation}
\label{eq:A b_solution_2_ass integral}
\begin{aligned}
\int_0^{\infty}d\rho~\rho^2\cdot e^{-(k+1)\rho}j_l(p\rho)
&=\frac{\sqrt{\pi}}{2^{l+1}}\frac{p^l}{(k+1)^{l+3}}\Gamma(l+3)\cdot
f\left(\frac{1}{2}(l+3),\frac{1}{2}(l+4),
l+\frac{3}{2};\frac{-p^2}{(k+1)^2}\right)  \\
f(a,b,c;z)&\equiv\frac{1}{\Gamma(c)}~{}_{2}F_{1}(a,b,c,z),
\end{aligned}
\end{equation}
where ${}_{2}F_{1}(a,b,c;z)$ is the Gauss hypergeometric function, and $f(a,b,c;z)$ is defined to be the regularized version of  ${}_{2}F_{1}(a,b,c,z)$ in a conventional way, see Refs. \cite{Abramowitz15:1972,Abramowitz6:1972} and recent article \cite{DiLella:2002ea}. As discussed in Sec. \ref{lensing}, we are especially interested in the non-relativistic domain, in this limit we have
\begin{equation}
\label{eq:A b_solution_2_ass integral_ass f}
\begin{aligned}
f\left(\frac{1}{2}(l+3),\frac{1}{2}(l+4),
l+\frac{3}{2};\frac{-p^2}{(k+1)^2}\right)
&\simeq\frac{1}{\Gamma(l+\frac{3}{2})}\left[
1-\frac{(l+3)(l+4)}{4(k+1)^2}
\frac{\Gamma(l+\frac{3}{2})}{\Gamma(l+\frac{5}{2})}p^2
+{\cal O}(p^4)\right]  \\
\end{aligned}
\end{equation}
Combing Eqs. \eqref{eq:A a solution_1}, \eqref{eq:A b_solution_2}, \eqref{eq:A b_solution_2_ass integral}, and \eqref{eq:A b_solution_2_ass integral_ass f}, we conclude
\begin{equation}
\label{eq:A a_solution_2}
\begin{aligned}
a_{plm}(t)
&=\eta_{lm}\frac{f_a\Delta R~e^{-\delta}}{2^{l+3}m_a^2}
\sqrt{\pi E_a}
\frac{\sin(E_a t)}{\sin(\frac{1}{2}E_a t_{\rm osc})}
\frac{\Gamma(l+3)}{\Gamma(l+\frac{3}{2})}\left(\frac{p}{m_a}\right)^l
H_l(p,\delta)  \\
&\simeq\eta_{lm}\frac{f_a\Delta R~e^{-\delta}}{2^{l+3}m_a^2}
\sqrt{\pi E_a}
\frac{\sin(E_a t)}{\sin(\frac{1}{2}E_a t_{\rm osc})}
\frac{\Gamma(l+3)}{\Gamma(l+\frac{3}{2})}\left(\frac{p}{m_a}\right)^l
H_l(0,\delta)\left[1+{\cal O}(p/m_a)^2
\right]
\end{aligned}
\end{equation}
where we define $H_l(p,\delta)$ to be the summation series
\begin{equation}
\label{eq:A b_solution_6_ass H}
\begin{aligned}
H_l(p,\delta)
&\equiv\sum_{n=0}^\infty\sum_{k=0}^n\frac{e^{-k\delta}}{2^n}
\frac{(n+1)!}{k!(n-k)!}\frac{(-1)^k}{(k+1)^{l+3}}
\Gamma(l+\frac{3}{2})f\left(\frac{1}{2}(l+3),\frac{1}{2}(l+4),
l+\frac{3}{2};\frac{-(p/m_a)^2}{(k+1)^2}\right).
\end{aligned}
\end{equation}
Then, the total radiation energy $E_{\rm rad}$ of the domain wall is obviously
\begin{equation}
\label{eq:A E excitation}
\begin{aligned}
E_{\rm rad}
&=\int d^3x\frac{1}{2}\chi
\left[-\frac{\partial^2}{\partial t^2}+L_2\right]\chi 
=\sum_{lm}\int d^3p\frac{1}{2}E_a |a_{plm}|^2  \\
&=\sum_{lm}\int_{m_a}^\infty dE_a\cdot 2\pi p~E_a^2|a_{plm}|^2.  \\
\end{aligned}
\end{equation}
More generally, assuming now the flux is produced within a ``cavity of radiation'' $V_{\rm rad}$, the density of radiation energy (per unit volume) is therefore $E_{\rm rad}/V_{\rm rad}$. Then the net flux $\Phi_{\rm rad}$ going through the boundary of the cavity is clearly
\begin{equation}
\label{eq:A flux_1}
\begin{aligned}
\frac{1}{S_{\rm rad}}\frac{d}{dE_a}\Phi_{\rm rad}
=\frac{p}{E_a^2}\frac{d}{dE_a}
\left(\frac{E_{\rm rad}}{V_{\rm rad}}\right) 
=\sum_{lm}\frac{2\pi p^2}{V_{\rm rad}}|a_{plm}|^2.
\end{aligned}
\end{equation}
Let $R_{\rm rad}\equiv\frac{V_{\rm rad}}{S_{\rm rad}}$ defines the effective size of cavity of radiation, we obtain
\begin{equation}
\label{eq:A flux_2}
\begin{aligned}
&\qquad\frac{d}{dE_a}\Phi_{\rm rad}
=\sum_{lm}\frac{\eta_{lm}^2}{R_{\rm rad}}
\frac{f_a^2\Delta R^2}{m_a^2}\frac{\pi^2e^{-2\delta}}{2^{2l+5}}
\left[\frac{\Gamma(l+3)}{\Gamma(l+\frac{3}{2})}\right]^2
\left[\frac{\sin(E_a t)}{\sin(\frac{1}{2}E_a t_{\rm osc})}\right]^2
E_a\left(\frac{p}{m_a}\right)^{2l+2}|H_l(p,\delta)|^2  \\
&\simeq\sum_{lm}\frac{\eta_{lm}^2}{R_{\rm rad}}
\frac{f_a^2\Delta R^2}{m_a^2}\frac{\pi^2e^{-2\delta}}{2^{2l+5}}
\left[\frac{\Gamma(l+3)}{\Gamma(l+\frac{3}{2})}\right]^2
\left[\frac{\sin(E_a t)}{\sin(\frac{1}{2}E_a t_{\rm osc})}\right]^2
E_a\left(\frac{p}{m_a}\right)^{2l+2}|H_l(0,\delta)|^2
+{\cal O}(p/m_a)^{2l+4}.
\end{aligned}
\end{equation}
A few comments should be made regarding to the magnitude of $R_{\rm rad}$. First, $V_{\rm rad}$ is defined as the cavity where radiation happens, so $R_{\rm rad}\simeq\Delta R$ in 1D case where thin-wall approximation is assumed. However, in 3D $R_{\rm rad}$ may extend to order of $R_0$ or even $m_a^{-1}$. More generally, it is reasonable to conjecture $R_{\rm rad}$ can depend on the angular momentum $l$. It is clear that to compute or even estimate the order of $R_{\rm rad}$ is very difficult. Thus, we should not bother the details of $R_{\rm rad}$, but rather treat it as a tunable normalization parameter and maybe absorb it into $\eta_{lm}$ if applicable.

We also express the spectra as a function of flux velocity
\begin{equation}
\label{eq:A flux v}
\begin{aligned}
&\qquad\frac{d}{dv_a}\Phi_{\rm rad}
=\sum_{lm}\frac{\eta_{lm}^2}{R_{\rm rad}}
\frac{f_a^2\Delta R^2}{m_a^3}\frac{\pi^2e^{-2\delta}}{2^{2l+5}}
\left[\frac{\Gamma(l+3)}{\Gamma(l+\frac{3}{2})}\right]^2
\left[\frac{\sin(E_a t)}{\sin(\frac{1}{2}E_a t_{\rm osc})}\right]^2
E_a^3\left(\frac{p}{m_a}\right)^{2l+3}|H_l(p,\delta)|^2  \\
&\simeq\sum_{lm}\frac{\eta_{lm}^2}{R_{\rm rad}}
\frac{f_a^2\Delta R^2}{m_a^3}\frac{\pi^2e^{-2\delta}}{2^{2l+5}}
\left[\frac{\Gamma(l+3)}{\Gamma(l+\frac{3}{2})}\right]^2
\left[\frac{\sin(E_a t)}{\sin(\frac{1}{2}E_a t_{\rm osc})}\right]^2
E_a^3\left(\frac{p}{m_a}\right)^{2l+3}|H_l(0,\delta)|^2
+{\cal O}(p/m_a)^{2l+5}.
\end{aligned}
\end{equation}
If the spherical symmetry is well preserved during most period of the annihilation of AQNs, then $\eta_{00}$ is the dominant term and Eq. \ref{eq:A flux v} can be simplified considerably.  We plot this (normalized) result in Figs. \ref{fig:4.2 3D_rhov}. One can see the final result is not very sensitive to the parameter $\delta$.  Such spectrum indicates an average energy $\langle E_a\rangle\simeq1.35m_a$ and an average velocity $\langle v_a\rangle\simeq0.6c$.  Comparing to the 1D case where $\langle E_a\rangle\simeq1.18m_a$ and $\langle v_a\rangle\simeq0.5c$ \cite{Fischer:2018niu}, we conclude that the general features of the spectrum in the relativistic regime for $v_a\geq 0.5 c$ is qualitatively consistent between 1D and 3D cases as anticipated in the original work \cite{Fischer:2018niu}. 
In particular, the difference between   these two cases  is about  $20\%$ for average velocity $\langle v_a\rangle$, 
and about $14\%$  for average energy $\langle E_a\rangle$. 
However, the spectra in the non-relativistic regime  $v_a\ll  c$ are dramatically different, see Fig. \ref{fig:1d3drhovfull}.

\begin{figure} [h]
	\centering
	\includegraphics[width=0.7\linewidth]{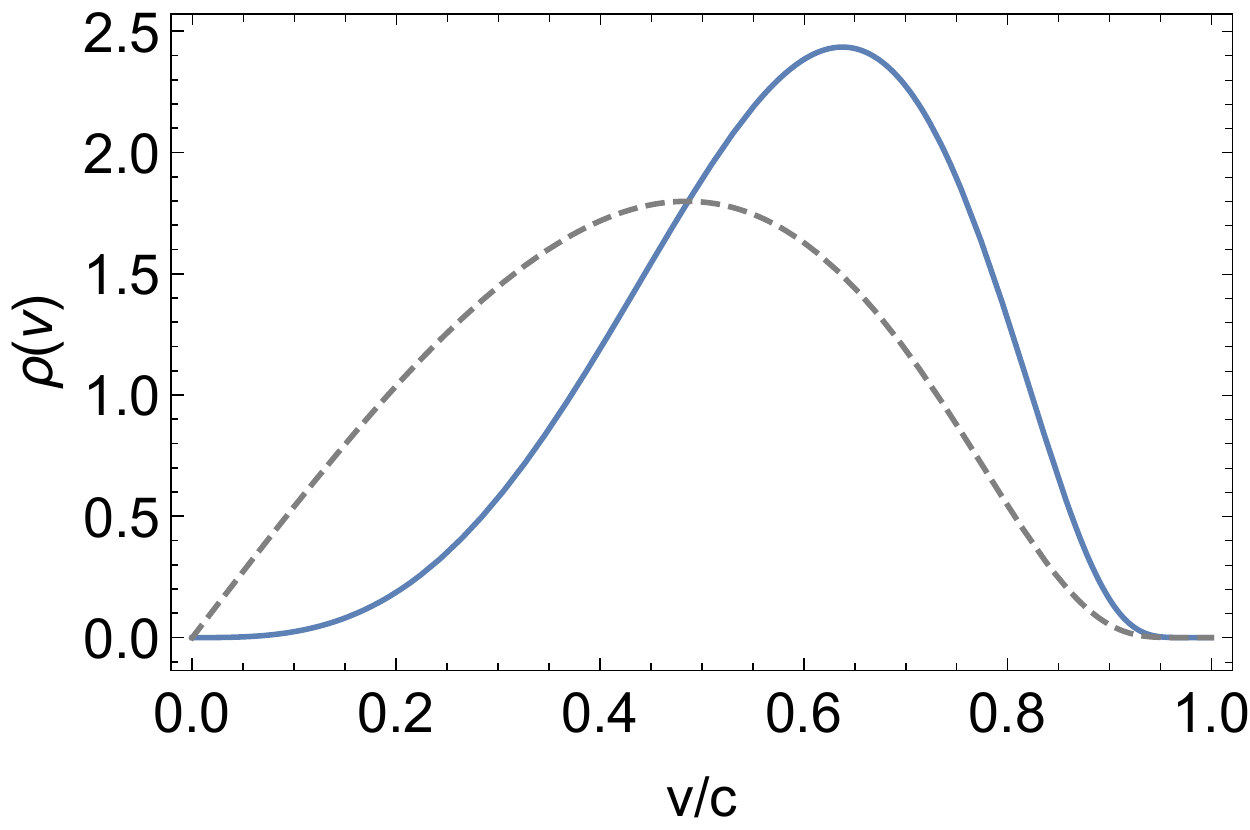}
	\caption{Axion flux spectrum: 1D versus 3D case. Here 1D case (gray dotted) computed in \cite{Fischer:2018niu} is compared with the 3D case (blue solid, $\delta=0$).}
	\label{fig:1d3drhovfull}
\end{figure}

 Lastly, it is instructive  to compare our approximate analytical solution \eqref{eq:A approximate DW solution} to the exact numerical solution. To do this, we plot the corresponding solutions in Fig. \ref{fig:A phisol} with $R_0$ chosen to be 0.01$m_a^{-1}$ for demonstration purpose. We find the exact numerical solution in general has a steeper growth, and approaches the outer  vacuum expectation value faster comparing to approximate solutions with $0\leq\delta\lesssim 1$. Beyond $\delta\lesssim1$ we find a best-fit solution at about $\delta=8$, but we should also note such solution badly violates the continuity and first-order differentiability at the transition zone $r\sim R_{\rm trans}$. We also plot $\delta=8$ in the flux spectrum Figs. \ref{fig:4.2 3D_rhov}, which again gives qualitatively consistent answer.

 As a final remark, one should not consider the spectrum with $\delta=8$ any better than other values of $\delta$ for two reasons. First, solution with $\delta=8$ is not physical for its bad violation of continuity and smoothness. Second, even an exact numerical solution gives no better quantitative prediction because the effective Lagrangian \eqref{eq:A L(phi)} is only a phenomenological model for qualitative analysis, see footnote \ref{footnote:A qualitative model}.

\begin{figure}
	\centering
	\includegraphics[width=0.7\linewidth]{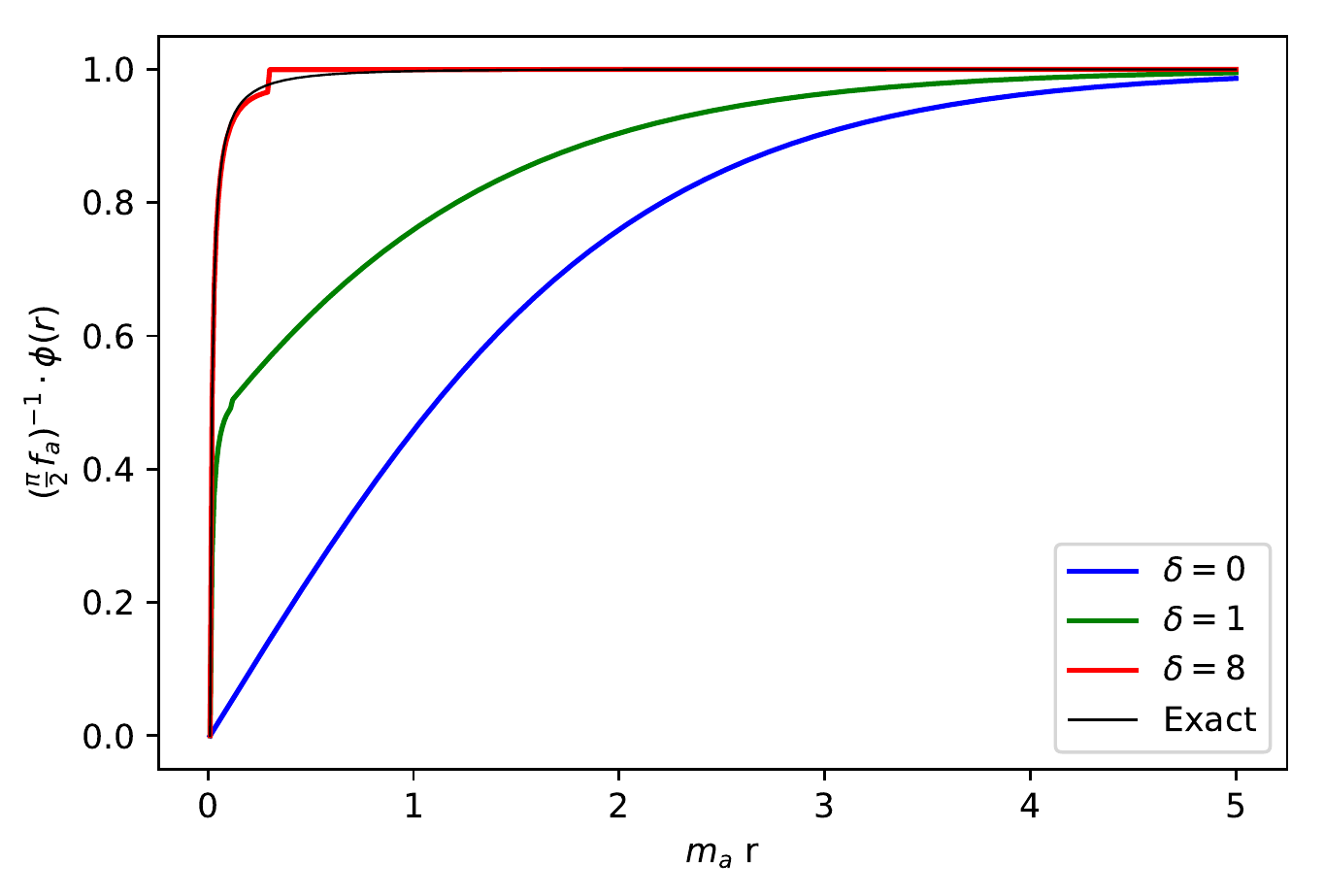}
	\caption{Approximate and exact numerical solutions to Eq. \eqref{eq:A equation of motion}. Here we choose $m_a R_0 = 0.01$.}
	\label{fig:A phisol}
\end{figure}

\section{About $K_{p,q}^{(l)}$}
\label{appendix:B About K}

This appendix is devoted to prove $K_{p,q}^{(l)}\delta(p-q)=\delta(p-q)$. Before we proceed to the proof, it is convenient to define two operators
\begin{equation}
\label{eq:B d}
d_l=\frac{d}{dr}+\frac{l+1}{r},\qquad
d_l^\dagger=-\frac{d}{dr}+\frac{l+1}{r}.
\end{equation}
And we will need some useful identities:
\begin{subequations}
\label{{eq:B d properties}}
\begin{equation}
\label{eq:B d properties_1}
d_ld_l^\dagger[rj_l(r)]=rj_l(r).
\end{equation}
\begin{equation}
\label{eq:B d properties_2}
d_l^\dagger[rj_l(r)]=rj_{l+1}(r)
\end{equation}
\begin{equation}
\label{eq:B d properties_3}
d_l^\dagger d_l=d_{l+1}d_{l+1}^\dagger
\end{equation}
\begin{equation}
\label{eq:B d properties_4}
\int_0^\infty [d_lA(r)]\cdot B(r)
=\int_0^\infty A(r)\cdot [d_{l}^\dagger B(r)]
\quad({\rm if~}A,B=0{\rm~~at~~}r=0,\infty)
\end{equation}
\end{subequations} 
Now we are able to prove by induction. First, the proof in case of $l=0$ is quite trivial:
\begin{equation}
\label{eq:B K_l=0}
\begin{aligned}
K_{p,q}^{(0)}
&=\lim_{L\rightarrow\infty}
\frac{\int_0^L dr~\sin(pr)\sin(qr)
\frac{1}{2}\left[
3\left(\frac{1}{v}\phi_{w,R_0}(r)\right)^2-1\right]}
{\int_0^L dr~\sin(pr)\sin(qr)}  \\
&=\lim_{L\rightarrow\infty}
\frac{\int_0^L dr~
[\cos(p-q)r-\cos(p+q)r]
\frac{1}{2}\left[
3\left(\frac{1}{v}\phi_{w,R_0}(r)\right)^2-1\right]}
{\int_0^L dr~[\cos(p-q)r-\cos(p+q)r]}. 
\end{aligned}
\end{equation}
Up to this point, we note that $K_{p,q}^{(0)}$ is always finite for any positive $p,q>0$. Now, if we multiply both sides by $\delta(p-q)$
\begin{equation}
\label{eq:B K delta_l=0}
\begin{aligned}
K_{p,q}^{(0)}\delta(p-q)
&=\delta(p-q)
\lim_{L\rightarrow\infty}
\frac{\int_0^L dr~
[1-\cos(p+q)r]
\frac{1}{2}\left[
3\left(\frac{1}{v}\phi_{w,R_0}(r)\right)^2-1\right]}
{\int_0^L dr~[1-\cos(p+q)r]}  \\
&=\delta(p-q),
\end{aligned}
\end{equation}
where we know the integral is quickly dominant by the the term $\int_0^Ldr\cdot1$, so that the fraction in the limit $L\rightarrow\infty$ gives trivial result.

Now, we want to show that $K_{p,q}^{(l)}\delta(p-q)=K_{p,q}^{(l+1)}\delta(p-q)$ for all $l=0,1,2...$. First, let us see that
\begin{equation}
\label{eq:B K_l!=0}
\begin{aligned}
K_{p,q}^{(l)}
&=\lim_{L\rightarrow\infty}\frac{\int_0^L dr\cdot
d_ld_l^\dagger[prj_l(pr)]\cdot
d_ld_l^\dagger [qrj_l(qr)]\cdot
\frac{1}{2}\left[
3\left(\frac{1}{v}
\phi_{w,R_0}(r)\right)^2-1\right]}
{\int_0^L dr\cdot 
d_ld_l^\dagger[prj_l(pr)]\cdot
d_ld_l^\dagger [qrj_l(qr)]}  \\
&=\lim_{L\rightarrow\infty}\frac{\int_0^L dr\cdot
prj_{l+1}(pr)\cdot
d_l^\dagger d_l[qrj_{l+1}(qr)]\cdot
\frac{1}{2}\left[
3\left(\frac{1}{v}
\phi_{w,R_0}(r)\right)^2-1\right]}
{\int_0^L dr\cdot 
prj_{l+1}(pr)\cdot
d_l^\dagger d_l[qrj_{l+1}(qr)}-  \\
&\qquad-\lim_{L\rightarrow\infty}
\frac{\int_0^L dr\cdot
prj_{l+1}(pr)\cdot qrj_{l+1}(qr)\cdot
\frac{1}{2}\frac{d}{dr}\left[
3\left(\frac{1}{v}
\phi_{w,R_0}(r)\right)^2-1\right]}
{\int_0^L dr\cdot 
prj_{l+1}(pr)\cdot
d_l^\dagger d_l[qrj_{l+1}(qr)]}  \\
&=K_{p,q}^{(l+1)}
-\lim_{L\rightarrow\infty}
\frac{\int_0^L dr\cdot r^2
j_{l+1}(pr)\cdot j_{l+1}(qr)\cdot
\frac{1}{2}\frac{d}{dr}\left[
3\left(\frac{1}{v}
\phi_{w,R_0}(r)\right)^2-1\right]}
{\int_0^L dr\cdot r^2
j_{l+1}(pr)\cdot
j_{l+1}(qr)},  \\
\end{aligned}
\end{equation}
where we have applied Eqs. \eqref{eq:B d properties_1} to \eqref{eq:B d properties_4} in the intermediate steps. Again, the integral is finite for any $p,q>0$. Note that in the last line if we set $p=q$, the second term must vanish. It is because the numerator is obviously finite, while the denominator tends to infinity in the large $L$ limit as indicated by Eq. \ref{eq:A orthogonality_J}. Thus, we conclude
\begin{equation}
\label{eq:B K delta_l!=0}
\begin{aligned}
K_{p,q}^{(l)}\delta(p-q)=\delta(p-q)
\end{aligned}
\end{equation}
for all $l=0,1,2,3...$ as expected.


\end{document}